\PassOptionsToPackage{table}{xcolor}
\documentclass[sigconf,nonacm]{acmart}

\usepackage{tabularx}
\usepackage{array}
\usepackage{makecell}
\usepackage{comment}
\usepackage{longtable}
\usepackage{booktabs}
\usepackage{multirow}
\usepackage{siunitx}
\usepackage{xcolor}
\usepackage{fvextra}

\definecolor{ourrow}{gray}{0.92}

\usepackage{tcolorbox}
\usepackage{fontawesome5}
\definecolor{lensSubject}{HTML}{F0534F}
\definecolor{lensPhysical}{HTML}{FFA000}
\definecolor{lensStylistic}{HTML}{44A047}
\definecolor{lensCreator}{HTML}{1E88E5}

\newtcbox{\lenschip}[1]{on line, arc=2pt, outer arc=2pt, boxrule=0pt, boxsep=0pt,
  left=2.5pt, right=2.5pt, top=1pt, bottom=1pt,
  colback=lens#1!16, colframe=lens#1!16}
\DeclareRobustCommand{\lenstag}[2]{\leavevmode\smash{\lenschip{#1}{%
  \color{lens#1!70!black}{\footnotesize\csname lensicon#1\endcsname}\hspace{2.5pt}%
  {\ttfamily\bfseries\fontsize{7.5pt}{9pt}\selectfont\vphantom{Ag}#2}}}}
\newcommand{\LensSubject}{\lenstag{Subject}{Subject}}
\newcommand{\LensPhysical}{\lenstag{Physical}{Physical Characteristics}}
\newcommand{\LensStylistic}{\lenstag{Stylistic}{Stylistic and Chronological}}
\newcommand{\LensCreator}{\lenstag{Creator}{Creator}}

\newcolumntype{Y}{>{\raggedright\arraybackslash}X}
\newcolumntype{C}[1]{>{\centering\arraybackslash}p{#1}}
\newcommand{\rowrule}{\arrayrulecolor{black!35}\specialrule{0.3pt}{1pt}{1.2pt}\arrayrulecolor{black}}

\AtBeginDocument{%
  }

\begin{document}

\title[MyMuse: A Personal Museum for Post-Visit Inquiry and Reflection]{MyMuse: A Personal Museum\\ for Post-Visit Inquiry and Reflection}

\author{Jiyeon Bae}
\affiliation{%
  \institution{Seoul National University}
  \city{Seoul}
  \country{Republic of Korea}}
\affiliation{%
  \institution{Harvard University}
  \city{Cambridge}
  \state{MA}
  \country{USA}}
\email{jiyeonb@snu.ac.kr}

\author{Chunggi Lee}
\affiliation{%
  \institution{Harvard University}
  \city{Cambridge}
  \state{MA}
  \country{USA}}
\email{chunggi\_lee@g.harvard.edu}

\author{Jinwook Seo}
\affiliation{%
  \institution{Seoul National University}
  \city{Seoul}
  \country{Republic of Korea}}
\email{jseo@snu.ac.kr}

\author{Hanspeter Pfister}
\affiliation{%
  \institution{Harvard University}
  \city{Cambridge}
  \state{MA}
  \country{USA}}
\email{pfister@seas.harvard.edu}

\renewcommand{\shortauthors}{Bae et al.}

\begin{abstract}
Artwork photographs may preserve what visitors saw without preserving why it mattered. Revisiting them can involve recalling earlier responses and developing new questions and interpretations as knowledge and personal experiences change. Yet recurring interests may remain difficult to recognize across individual exchanges. Informed by a formative study with ten visitors, we present MyMuse, a post-visit system that connects inquiry about individual artworks with reflection across visitors' accumulated records. In its Personal Museum, visitors revisit artworks, ask questions, and record interpretations that inform subsequent system responses. Its interactive Narrative Report reorganizes the same records across artworks through four interpretive lenses. In two within-subject studies with 16 participants, direct comparisons favored MyMuse for vivid revisitation. The report received higher ratings for recurring-interest awareness, cross-artwork connections, and unexpected discovery than a per-artwork presentation of the same records. These findings suggest that examining accumulated records can support reflection on recurring interests and relationships among visitors' responses across artworks.
\end{abstract}




\begin{teaserfigure}
    \centering
    \includegraphics[width=\textwidth]{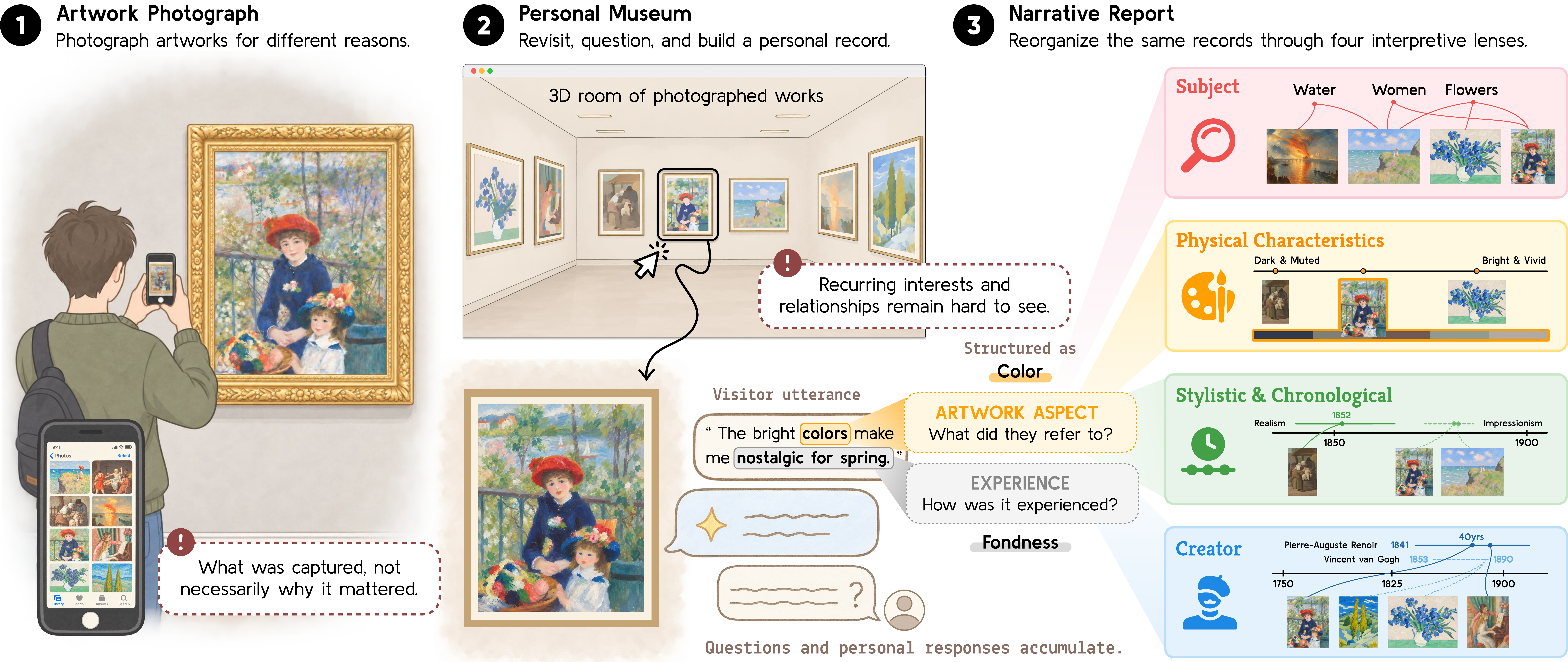}
    \caption{MyMuse is a web application that supports post-visit inquiry and reflection around visitors' artwork photographs. Visitors revisit photographed works in a Personal Museum, where their questions and reflections accumulate around individual artworks and are annotated by artwork aspect and experiential trigger. The Narrative Report reorganizes the same records through four interpretive lenses so visitors can examine recurring interests and relationships across works.}
    \Description{A three-stage overview arranged from left to right. First, a visitor photographs an artwork, with a phone gallery showing collected artwork photographs. Second, the Personal Museum displays photographed works in a 3D room and links a selected artwork to an ongoing conversation. An example visitor utterance about bright colors and nostalgia for spring is highlighted to illustrate the artwork-aspect annotation Color and the experiential annotation Fondness. Third, four Narrative Report views reorganize the collection through the Subject, Physical Characteristics, Stylistic and Chronological, and Creator lenses, illustrated with subject links, a color-based arrangement, and timelines.}
    \label{fig:teaser}
\end{teaserfigure}

\maketitle

\section{Introduction}
\label{sec:introduction}

Art encounters can shape what visitors notice, remember, and find meaningful. This engagement can continue beyond the museum visit. Visitors may photograph artworks that draw their attention and later return to these images to remember or reconsider the experience. Photo-taking can direct attention toward selected aspects of an experience \cite{diehl16}. As visitors learn more about an artwork and its context, they may notice different aspects of the work or reconsider their interpretation \cite{chatterjee14,morse23}. Research on revisiting stored traces also suggests that engagement can extend beyond the original encounter \cite{chen23,wang24,shapiro18,yin25}. Together, these studies provide a basis for examining how artwork photographs can support inquiry and reinterpretation after a visit \cite{vavoula09}.

However, a photograph may preserve the appearance of an artwork without preserving the visitor's response to it. Returning to the image may therefore leave unclear why the work drew their attention or what they wanted to explore. Research on personal records cautions that more complete capture does not necessarily support better remembering \cite{sellen10}. Prior work has explored connecting material collected during museum visits with subsequent learning activities \cite{vavoula09}, but access to this material does not by itself make visitors' questions and interpretations available for later use. Across a collection, similarities among photographed artworks may provide limited evidence of relationships among visitors' own responses. Even when questions and interpretations are recorded, connections among them may remain difficult to recognize across separate artwork entries. We examine how a post-visit system can use accumulated personal records to inform subsequent inquiry and help visitors examine recurring interests and relationships across works.

To understand what visitors need for post-visit inquiry and reflection, we conducted a qualitative formative study with ten museum and gallery visitors using 100 photographs from their own visits. Participants described difficulty recalling why they had photographed particular works. Their accounts also indicated how later knowledge could prompt reconsideration and how accumulated photographs could invite reflection on recurring interests. These findings informed the design of \textbf{MyMuse}, a post-visit personal museum in which visitors can add to and revisit records associated with their artwork photographs (Figure~\ref{fig:teaser}). At the individual-work level, visitors can review earlier entries, ask further questions, articulate responses, and explore related artworks. MyMuse uses prior records to inform subsequent responses while keeping visitor-authored content distinct from retrieved context and system-generated text. At the collection level, a \textbf{Narrative Report} organizes accumulated records through interpretive lenses adapted from \emph{Cataloging Cultural Objects} (CCO) \cite{baca06}. Its composition is computed deterministically from available evidence, while language generation phrases the selected observations. Drawing on narrative visualization principles \cite{segel10}, the report presents a guided account of recurring aspects and relationships that visitors can examine alongside the supporting records.

To examine whether MyMuse supports these two levels of engagement, we conducted two controlled within-subject studies in a single session with 16 participants using their own artwork photographs. In direct comparisons with a baseline using a separate, randomly assigned subset of each participant's artwork photographs, 13 of 16 participants favored MyMuse for vividly revisiting prior museum experiences, and 12 favored it for noticing cross-work connections. At the collection level, the Narrative Report received higher ratings for recurring-interest awareness, cross-artwork connections, and unexpected discovery than a per-artwork report derived from the same interaction records. Perceived fidelity did not significantly differ. Qualitative accounts indicated that reflection across records included articulating familiar interests more precisely. These findings suggest that organizing visitors' own records can complement artwork revisitation by supporting reflection on what connects their responses across works.

This work makes three contributions. First, we provide a qualitative account of visitors' difficulties and needs when revisiting artwork photographs, including how later knowledge and accumulated collections can prompt reinterpretation and reflection. Second, we present \textbf{MyMuse}, a post-visit personal museum that reuses the questions and interpretations visitors record while exploring individual artworks, both as context for subsequent inquiry and as evidence in the Narrative Report, which organizes these records visually across works so that visitors can reflect on what they responded to and how. Third, we provide evidence from a controlled evaluation concerning visitors' experiences of vivid revisitation, cross-work connections, and awareness and articulation of recurring interests.
\section{Related Work}

\subsection{Photography as Selective Attention and Personal Record}
\label{sec:rw-photography}

Photographing artworks can influence what visitors attend to and later remember.
Henkel found that photographing whole museum objects impaired later memory relative to observation, while photographing a selected detail did not \cite{henkel14}.
Volitional photography can also direct visual attention toward aspects that people choose to capture \cite{diehl16, barasch17}.
These findings suggest that capture involves selective attention rather than passive storage.

Visitors' reasons for photographing a work may not be apparent from the image alone.
Museum visitors photograph artworks for different reasons, including memory, research, inspiration, sharing, and self-expression \cite{lambert17}.
Research on memory-oriented records similarly emphasizes the value of selective rather than exhaustive capture \cite{sellen10,petrelli09}.
Niforatos et al.\ further found that manually captured photographs were perceived as more meaningful than automatically captured ones \cite{niforatos17}.
Yet intentional selection does not necessarily leave its rationale explicit: Petrelli et al.\ found that carefully selected personal cues often contained little annotation explaining why they had been chosen \cite{petrelli09}. In Myartspace, students revisiting museum photographs struggled to identify some images and relate them to accompanying audio and text comments that were not linked to the photographs \cite{vavoula09}.
An artwork photograph can therefore preserve \textit{what} was selected without preserving \textit{why} it mattered.

\subsection{Context in Artwork Interpretation}
\label{sec:rw-meaning}

Museum encounters are shaped by what visitors bring to them and by the contexts in which they occur \cite{pearce16,fosh15}.
Using the Contextual Model of Learning, Falk and Storksdieck found that interacting personal, social, and physical factors influenced visitor learning, while no single factor adequately explained outcomes across visitors \cite{falk05}.
In art viewing, Brieber et al.\ found that artworks encountered in a museum were evaluated more positively and remembered better than computer-presented reproductions, with museum viewers also using spatial-layout cues during retrieval \cite{brieber15}.

Meaning can also remain open beyond the initial encounter.
Morse et al.\ found that meaningful museum experiences could develop before, during, or after a visit, involve multiple interacting sources of meaning, and remain fundamentally subjective at the individual level \cite{morse23}.
Chatterjee and Vartanian review evidence that contextual information such as titles, authenticity, stylistic knowledge, and historical or cultural context can influence aesthetic responses \cite{chatterjee14}.
In a sculpture-garden study, Fosh et al.\ invited visitors to interpret works before receiving official explanations and found that their interpretations sometimes diverged from those provided \cite{fosh13}.
Together, this work suggests that what becomes meaningful in an artwork is not determined by the image alone \cite{walton70} and can be reconsidered as additional context becomes available \cite{bullot13}.

\subsection{Revisiting and Organizing Personal Records}
\label{sec:rw-records}

Personal records can support reflection when they remain accessible for later revisiting and organization \cite{odom14,elsden16,xu24}.
Work on personal informatics shows that collection, integration, and reflection are interdependent stages of working with personal data \cite{li10}.
Similar challenges appear in creative practice, where accumulated material can become difficult to revisit or understand as a whole \cite{dalsgaard23,turco25}.

Prior systems offer ways to extend earlier records and organize personal collections.
In Echo, users add reflections to earlier records of everyday experiences and review their changing reactions to the same events \cite{isaacs13}.
Personal curation allows people to collect and reorganize material around their own interests \cite{shapiro18,weilenmann13,zhao14}.
TravelGalleria uses spatial organization to support remembering and reflection across personal photographs \cite{yin25}, while SemanticCollage supports exploring relationships and obtaining overviews across visual collections \cite{koch20}.

Other systems use contextual information to support inquiry about personal media and artworks.
OmniQuery integrates context across personal media to support personal question answering \cite{li25}, while VirtuWander uses visitor inquiries and tour context to provide guidance, artwork comparisons, and tour summaries in a virtual museum \cite{wang24virtuwander}.
ArtRAG uses structured cultural, historical, stylistic, and thematic knowledge to support grounded, multi-perspective artwork explanations \cite{wang25}.

Narrative presentations have also been used to organize recorded information and connect guided reading with further exploration.
Martinez-Maldonado et al.\ used layered storytelling to support reflection on recorded learning activities, combining focused views with contextual annotations \cite{maldonado20}.
Gratzl et al.\ linked visual stories to their underlying analysis states, allowing readers to move from a guided presentation into further exploration \cite{gratzl16}.
Automated data-storytelling systems such as DataShot and Calliope have further explored how salient facts can be selected and composed into structured narratives from data \cite{shi21,wang20}.

MyMuse differs from the systems reviewed above in the material it works with and in how that material is reused.
Myartspace, Echo, and TravelGalleria support revisiting records of past experiences. VirtuWander and ArtRAG support inquiry about artworks encountered in a virtual tour or presented as images, and OmniQuery answers questions about personal media.
MyMuse applies grounded artwork inquiry to the photographs visitors took during their own museum visits, retains the questions and interpretations they add as verbatim records, and uses these records to inform later responses about other works.
At the collection level, it reorganizes the same records to support reflection on recurring interests and relationships across works.

\section{Qualitative Formative Study}
\label{sec:formative}

Artwork photographs may not preserve why a work mattered, how additional context influenced its interpretation, or how it relates to other collected works. We conducted a formative study using participants' own artwork photographs to examine what they recalled about past museum encounters, what they could no longer recover, and how they wanted to use these records. Their accounts informed MyMuse's support for revisiting individual artworks and examining relationships across a collection.

\subsection{Study Setup}
\label{sec:formative-setup}

\textbf{Participants and procedure.} We recruited ten participants ($N=10$; 9 women, 1 man; ages 21--28) who had visited a museum or gallery within the previous year. Each participant submitted ten artwork photographs from their own visits, yielding 100 photographs. We used photographs from past visits, rather than from a visit arranged for the study, so that the photographs had been taken for participants' own purposes and so that we could examine what participants could and could not recover from them some time after a visit. Each 60-minute session combined a semi-structured interview about museum-going, photography, information-seeking, and revisitation practices with a photo-elicitation interview using the participant's photographs. For each photograph, we asked why it had been taken, what remained memorable, what had been lost, and whether participants had acquired new knowledge about the work or changed how they intended to use the photograph. We also asked how they would like a system for revisiting artwork photographs, recording reflections, and asking questions to be organized. The study was approved by our institution's IRB, and all participants provided informed consent.

\textbf{Analysis.} We analyzed 538 question--answer units using reflexive thematic analysis \cite{braun06}, producing four themes. Separately, we used the nine chapter-level scopes of \emph{Cataloging Cultural Objects} (CCO) \cite{baca06} to characterize the range of artwork-related interests expressed by participants. For this analysis, two researchers coded 90 participant--chapter judgments with high agreement ($\kappa=.81$; 92.2\%). The coverage analysis informed the interpretive lenses described in Section~\ref{sec:design}. Interview topics, illustrative code--theme connections, and details of the CCO analysis are in Appendix~\ref{app:formative}.

\subsection{Findings}
\label{sec:formative-findings}

We identified four themes concerning what participants could recover from their photographs, how they reconsidered artworks, and what they wanted to understand across their collections.

\paragraph{F1. A photograph could preserve the artwork without preserving the reason for selecting it.}
Participants often remembered choosing a work without remembering what had prompted the choice. P07 described the difficulty directly: \textit{``I only took a few photos at the Banksy Museum. If I took this one, there was definitely a reason, but I don't remember it.''} Participants also reported losing titles, artists, and locations, so identifying the work was sometimes a prerequisite for reconsidering their earlier response. Prior work shows that photography is selective and that visitors capture works for different reasons \cite{diehl16,lambert17}. These accounts suggest that photographs alone may provide insufficient context for visitors to recall what drew their attention to a work.

\paragraph{F2. Responses to artworks drew on personal knowledge, roles, and experiences.}
\looseness=-1
What participants noticed in a work was shaped by their knowledge, roles, and experiences, consistent with prior work on personal context in museum experiences and art appreciation \cite{falk05,chatterjee14}. P06, who worked as an assistant curator, described differences between visiting for personal and professional purposes: \textit{``As a visitor, I focus on which paintings I like. At work, I look more at why a particular work was selected and how visitors experience the exhibition.''} P05 associated an installation of brightly colored pills with a period of taking medication and imagined how that experience might have felt with pills like those in the work. These accounts suggest that understanding participants' responses may require personal context that is not apparent from information about the artwork itself.

\paragraph{F3. New knowledge and related works could change which features visitors attended to.}
\looseness=-1
With additional contextual knowledge, participants described noticing different aspects of familiar artworks. After studying Korean art history, P06 recalled shifting attention from the colors and composition of Yoo Young-kuk's work toward how landscapes were translated into geometric forms. P07 described recognizing previously unnoticed qualities after learning why Cézanne was distinctive: \textit{``I see, he does seem to paint that way.''} Participants also described returning to earlier works after encountering related artworks. P08 described how encountering works by the same artist or on similar subjects could change an interpretation formed earlier, and P01 searched earlier museum photographs when a new work seemed familiar, wanting to establish how the two differed. The context participants found relevant varied, from historical period to artistic technique. Visitors may therefore return to a familiar work with questions informed by later learning and encounters with other artworks.

\paragraph{F4. Recurring subjects across a collection did not necessarily indicate personal interests.}
Participants described their photographs as a collection they wanted to examine, grouping works by visit or by personally meaningful theme and asking whether recurring subjects reflected broader interests. Their photographs also reflected purposes such as documentation and possible future use. P09 photographed works as potential material for writing and program planning, and cautioned that works captured because they seemed important to document could \textit{``contaminate''} an inferred taste profile. P07 suggested that repeatedly asking about an artist or a historical period could indicate an interest that the photographs alone would not make apparent. These accounts suggest that visitors may use their reasons for taking photographs and the questions they ask repeatedly to interpret recurring subjects across their collections.
\section{Design Considerations}
\label{sec:design}

We translated the formative findings and follow-up design feedback into four design goals for MyMuse. These goals address how to preserve visitors' contributions and support inquiry and reflection within and across artwork records.

\paragraph{DG1. Keep visitors' own contributions identifiable within artwork records.}
To retain the personal context of visitors' responses (F1, F2), MyMuse stores their questions and reflections verbatim alongside each artwork photograph. These contributions remain distinct from catalogue information and system-generated responses, so visitors can identify what they contributed and revisit their own words during subsequent inquiry.

\paragraph{DG2. Keep artwork records open to further inquiry.}
Participants described how new knowledge and related artworks could prompt reconsideration (F3). We therefore designed records to accommodate additional questions and interpretations while keeping earlier entries available. Visitors can consult contextual information, explore related artworks, and add responses arising from that exploration. To offer directions for inquiry, we used \emph{Cataloging Cultural Objects} (CCO) \cite{baca06} as an external descriptive reference. We selected four artwork-centered lenses based on their coverage in the formative data and suitability for guiding renewed attention: \LensCreator{}, \LensPhysical{}, \LensStylistic{}, and \LensSubject{}. These lenses are optional, and visitors can also record personal associations and concerns beyond their scope.{\emergencystretch=6em\par}

\paragraph{DG3. Ground collection-level patterns in visitor-authored records.}
Participants wanted to examine recurring interests, while noting that photographing a work did not necessarily indicate preference (F4). The Narrative Report therefore organizes visitor-authored records across artworks and presents recurring aspects and relationships alongside the supporting entries. Visitors can inspect this evidence and consider whether the reported connections reflect their interests.

\paragraph{DG4. Connect individual-work inquiry with collection-level reflection.}
Participants' interest in revisiting a personally meaningful collection (F4) motivated support for both examining relationships across works and returning to particular artworks. We designed complementary views that visitors can move between. From a collection-level pattern, visitors can return to the relevant artworks and records to examine the context of their contributions and continue their inquiry.
\begin{figure*}[t]
    \centering
    \includegraphics[width=\textwidth]{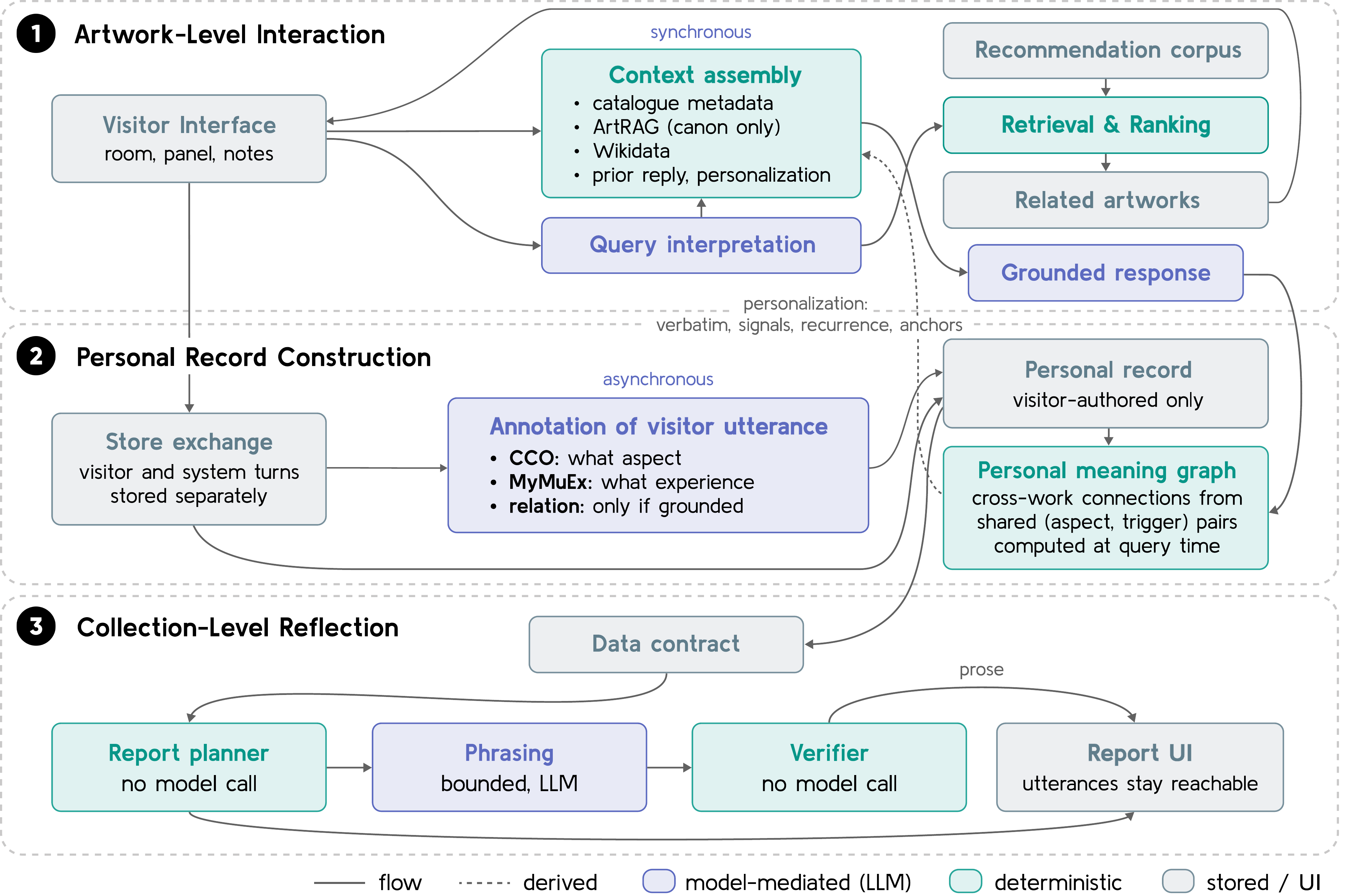}
    \caption{MyMuse system architecture. Depending on the visitor's request, MyMuse generates a grounded response, retrieves related artworks, or performs both operations. Visitor and system turns are stored separately. Only visitor-authored text is annotated and incorporated into the personal meaning record. Cross-work connections in the Personal Meaning Graph (PMG) are computed at query time from shared annotated (aspect, trigger) pairs and are not stored as persistent edges. For collection-level reports, a deterministic planner supplies the report structure and fallback wording directly to the report interface. Language-model output is limited to phrasing selected observations and is verified before use.}
    \Description{A system architecture diagram arranged in three horizontal bands. The top band shows artwork-level interaction, including query interpretation, context assembly, grounded response generation, and related-work retrieval. The middle band shows separate storage of visitor and system turns, asynchronous annotation of visitor utterances with CCO elements, MyMuEx triggers, and relations supported by the text, and query-time computation of cross-work PMG connections. The bottom band shows participant-level records organized through a data contract and provided to a deterministic report planner. Report structure and fallback wording are passed directly from the planner to the report UI. A separate path passes selected content through language-model phrasing and verification before the generated wording is used. The report UI retains access to the supporting visitor utterances.}
    \label{fig:architecture}
\end{figure*}

\section{MyMuse: System Design}
\label{sec:system}

MyMuse supports post-visit engagement with artwork photographs through personal records that visitors can revisit and extend. At the artwork level, visitors can review a work, ask questions, add reflections, and explore related artworks. Their contributions are stored verbatim and used to inform subsequent inquiry. At the collection level, a Narrative Report reorganizes accumulated records to present recurring aspects and relationships across works. The following sections describe how MyMuse presents these records, uses them during interaction, and organizes them for collection-level reflection. Figure~\ref{fig:architecture} summarizes the system architecture, including artwork-level interaction, personal record construction, and collection-level reflection.

\begin{figure*}[t]
    \centering
    \includegraphics[width=\textwidth]{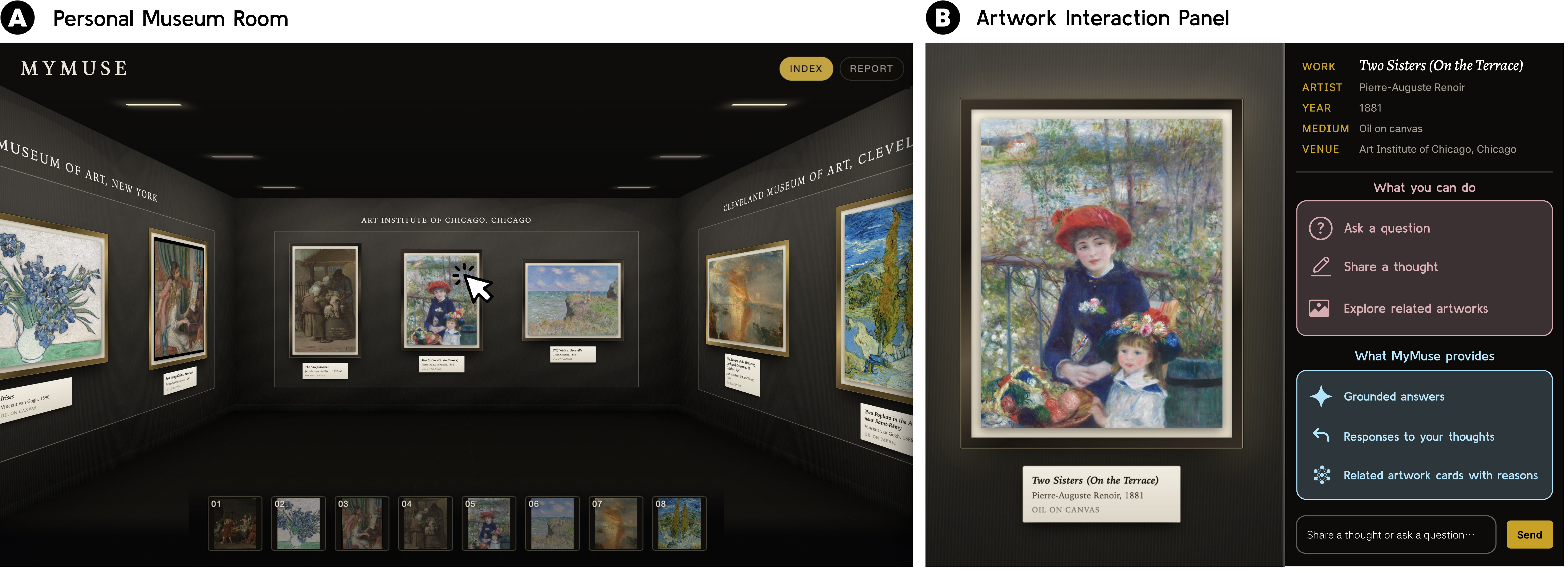}
    \caption{\textbf{MyMuse Personal Museum interface.} (A) The Personal Museum presents the visitor's photographed artworks in a gallery-like room for revisiting their collection. Artworks are grouped by exhibition, with a thumbnail strip for navigation. (B) Selecting an artwork opens an interaction panel displaying its image and metadata. Visitors can ask questions, share thoughts, and explore related artworks. MyMuse provides grounded answers, responses to visitor reflections, and related-artwork cards with explanations.}
    \Description{Two interface views labeled A and B. View A shows framed artworks arranged on three walls of a dark gallery-like room, grouped by exhibition, with eight artwork thumbnails along the bottom. A cursor indicates selection of Renoir's Two Sisters (On the Terrace). View B shows an enlarged image of the selected artwork beside its metadata and labeled summaries of supported visitor actions and system response types. A text input and Send button appear at the bottom of the interaction panel.}
    \label{fig:personal-museum}
\end{figure*}

\subsection{Returning to a Personal Museum Collection}
\label{sec:personal-museum}

MyMuse presents visitors' artwork photographs in a gallery-like room for returning to individual works and their associated records (Figure~\ref{fig:personal-museum}). The choice of a spatial representation was informed by participants' interest in revisiting their photographs as a personally meaningful collection (F4). Follow-up design feedback indicated varied preferences: some participants favored 2D views for scanning larger collections, while others valued a spatial representation for giving their photographs the feeling of ``my own museum.'' Prior work has similarly used gallery environments to support revisiting personal media in virtual space \cite{yin25}. These accounts informed our use of a gallery-like room alongside the Narrative Report, which supports comparing works and examining relationships across the collection.

MyMuse automatically groups photographs of works encountered in the same exhibition on the same wall. This arrangement retains their association with the exhibition the visitor attended without reproducing its original spatial layout. Each photograph is linked to a catalogue record with verified metadata, including title, artist, date, medium, and collection when available (Section~\ref{sec:implementation}). Selecting a work opens its interaction panel, where visitors can review existing records, ask questions, add personal observations, and explore related artworks.

\subsection{Grounded Artwork Inquiry}
\label{sec:grounded-inquiry}

MyMuse constructs responses from the artwork's catalogue record, external context, and selected prior records. The catalogue record is supplied as verified metadata together with its source and confidence level. The prompt instructs the model to ground factual claims about the work in this record, never to contradict or replace its title, artist, date, medium, or movement, and to keep qualifiers such as attributions and approximate dates. The model may add art-historical context from its own knowledge when it is consistent with the record, without presenting inferences as museum-record facts. MyMuse adds artist information from Wikidata, including movement, nationality, occupation, and life dates. When the artwork satisfies ArtRAG's medium and regional constraints and its graph contains the artist, MyMuse also adds structured art-historical context from ArtRAG \cite{wang25}, retrieved from the catalogue record and reranked against the visitor's photograph. The response context may also include MyMuse's previous response about the current work and a constrained selection from the visitor's prior records, which the prompt identifies as the visitor's perspective rather than as facts about the artwork. These sources are kept separate during context assembly so that catalogue information, retrieved context, system responses, and visitor-authored contributions remain distinguishable. Appendix~\ref{app:example-exchange} shows an exchange from the study with the sources used for each response.

\begin{figure*}[t]
    \centering
    \includegraphics[width=\linewidth]{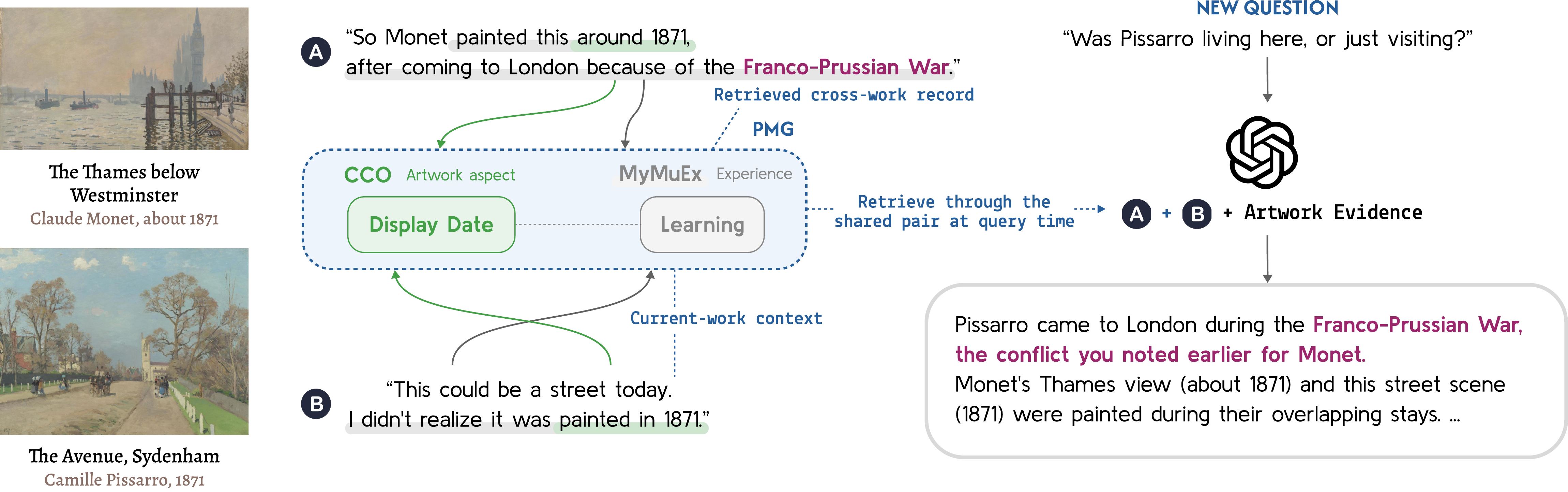}
    \caption{Constructed example illustrating how MyMuse uses the Personal Meaning Graph (PMG) to retrieve an earlier record for a subsequent response. Records~A and~B concern different artworks viewed by the same visitor and share the CCO aspect \textit{Display Date} and the MyMuEx trigger \textit{Learning}. At query time, MyMuse retrieves record~A through this shared annotation pair and includes its original wording alongside current-work record~B, artwork evidence, and the new question in the response context. The highlighted passage refers back to the visitor's earlier observation about Monet. Only annotations relevant to this connection are shown.}
    \label{fig:pmg-example}
    \Description{Visitor records about Monet's The Thames below Westminster and Pissarro's The Avenue, Sydenham appear beside a PMG showing their shared Display Date and Learning annotations. An arrow indicates retrieval of record A. The new question, retrieved record A, current-work record B, and artwork evidence are provided as inputs to response generation. The highlighted passage in the answer recalls the visitor's earlier reference to Monet and the Franco-Prussian War.}
\end{figure*}

\subsection{From Conversation to a Personal Meaning Record}

MyMuse stores visitor utterances verbatim and annotates them to support retrieval and comparison during later inquiry. System responses and derived annotations are stored separately from the original utterances. Annotation runs asynchronously after the visitor-facing response has been returned, allowing the interaction to proceed while the personal meaning record is updated.

The annotations use two published vocabularies. CCO \cite{baca06} describes which aspect of an artwork an utterance concerns, while MyMuEx \cite{morse23} describes how an experience becomes meaningful. From CCO's 123 elements, we designate 21 work-image elements as the primary visitor-addressable set. For subsequent inquiry and collection-level reflection, we organize these annotations through the four lenses described in Section~\ref{sec:design}. MyMuEx provides 23 experiential triggers (e.g., \emph{Learning}, \emph{Aesthetics}, and \emph{Inspiration}), with valence represented separately and temporality and scope describing when and how broadly an effect operates. For experiential annotation, MyMuse uses the 21 triggers identified in our formative study (Appendix~\ref{app:formative-mymuex}). The two vocabularies allow the record to distinguish the artwork aspect discussed from the visitor's experience of it.

CCO and MyMuEx are annotated independently. MyMuse records a relation between them only when the visitor's utterance supports that relation. Co-occurrence alone is insufficient. Each annotation retains its supporting span from the original utterance. Only visitor-authored text serves as evidence about the visitor, so information introduced in a system response does not become evidence of a personal interest or interpretation.

\subsection{Using Prior Records to Contextualize Inquiry}

MyMuse selects relevant prior records to inform a new request. It derives a \emph{Personal Meaning Graph (PMG)} from the accumulated annotations, connecting records across artworks when they share an annotated CCO aspect and MyMuEx trigger. These connections are computed at query time from the underlying records. They are used to retrieve relevant context, without being stored as persistent claims about the visitor's preferences or traits.

For each request, MyMuse draws on four forms of context. \emph{Verbatim recall} retrieves relevant remarks about the current work, while \emph{within-work signals} identify aspects revisited within that work. \emph{Cross-work recurrence} identifies aspects discussed across multiple works, and \emph{cross-work anchors} retrieve relevant remarks from other artworks connected through a shared (aspect, trigger) pair. These sources allow a response to refer to earlier contributions that bear on the current inquiry. Figure~\ref{fig:pmg-example} illustrates how PMG connections support retrieval of an earlier record from another artwork to inform a subsequent response.

\subsection{Exploring Related Works}

MyMuse provides related artworks for visitors to compare with the current work and pursue further questions. It retrieves these artworks from a corpus of approximately 12,800 works assembled from six sources: the Cleveland Museum of Art, the Metropolitan Museum of Art, Tate, the Art Institute of Chicago, Wikimedia Commons via Wikidata, and Google Arts \& Culture. This corpus is separate from the visitor's own photographs, and each item retains a reference to its source record.

Retrieval depends on the relationship requested by the visitor. Artist requests use explicit artist matching, supplemented by a live search of the Metropolitan Museum of Art's collection, while subject requests use controlled catalogue terms. For visual dimensions such as color, composition, or mood, MyMuse compares CLIP embeddings \cite{radford2021} of the artwork, cropped from the visitor's photograph where possible, with those of corpus works using dimension-specific textual probes. When no dimension is specified, retrieval uses overall embedding similarity. These methods support exploration through both catalogue relationships and visually derived similarities.

\begin{figure*}[t]
    \centering
    \includegraphics[width=\textwidth]{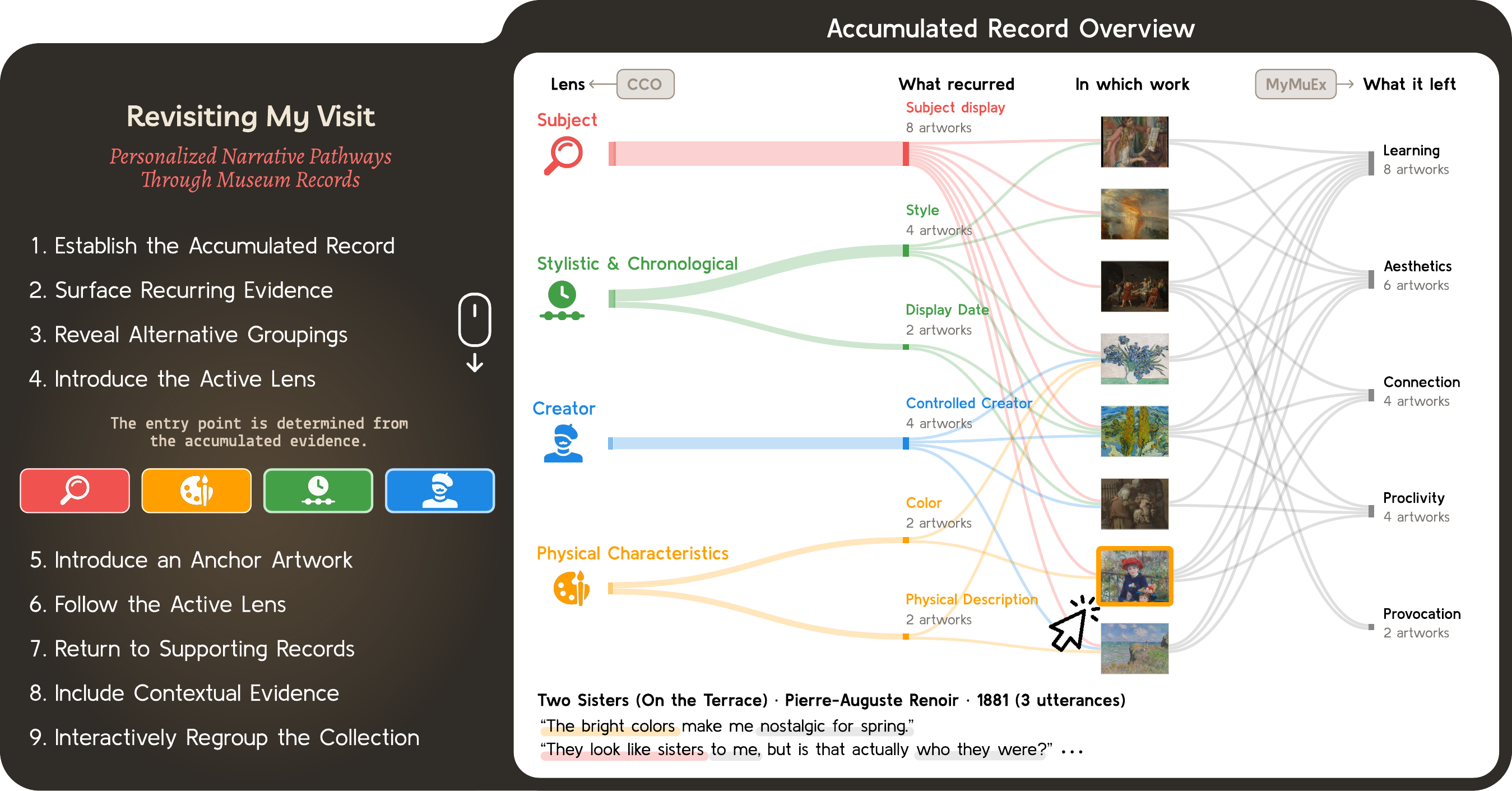}
    \caption{\textbf{MyMuse Narrative Report interface.} Left: the narrative structure used across four interpretive lenses, progressing from guided review of accumulated visitor records to interactive regrouping of the collection. MyMuse selects an entry point from these records and includes evidence-dependent stages when their required evidence is available. Right: the recurring-evidence view connects annotated CCO artwork aspects, artworks, and MyMuEx experiential triggers. Selecting an artwork displays supporting utterance excerpts below the diagram so visitors can examine the records underlying the annotations.}
    \Description{A two-panel interface figure. The left panel lists nine narrative stages and includes controls for four interpretive lenses: Subject, Physical Characteristics, Stylistic and Chronological, and Creator. The right panel shows a flow diagram connecting these lenses to recurring CCO artwork-aspect annotations, artwork thumbnails, and MyMuEx experiential triggers. Counts indicate the number of distinct artworks associated with each annotation. Renoir's Two Sisters (On the Terrace) is highlighted, with two supporting visitor-utterance excerpts displayed beneath the diagram.}
    \label{fig:narrative-report}
\end{figure*}

\subsection{Collection-Level Narrative Reflection}
\label{sec:report}

The Narrative Report reorganizes accumulated records across artworks so visitors can examine recurring aspects and relationships within their collection (Figure~\ref{fig:narrative-report}). We use narrative visualization to introduce these relationships progressively, linking selected observations to their supporting records. A deterministic planner selects the entry point, supporting evidence, and organization. Language generation then phrases the selected observations.

\textbf{Guided reading and interactive reorganization.} MyMuse presents the report through a scrollytelling interface. Artwork plates remain visible as their arrangement changes through the narrative, allowing visitors to follow the same works across different organizations. Visitors can inspect the utterances supporting the reported relationships. After the guided sequence, they can reorganize the collection themselves, following the martini-glass structure described by Segel and Heer \cite{segel10}. This transition allows visitors to examine the presented relationships and then explore other arrangements of their records.

\textbf{Preparing the evidence.} For each participant, the planner receives a structured record containing visitor-authored contributions, artwork metadata, annotations, supporting evidence spans, contextual records, and provenance. These are the same visitor-authored records used by the artwork-level system. System-generated replies are excluded as evidence about the visitor. For each of the four lenses, support is measured by the number of distinct artworks carrying evidence from the corresponding CCO chapter. Repeated discussion of one work therefore does not increase the number of works supporting a collection-level pattern. An aspect must occur across at least two distinct artworks to qualify as a collection-level recurrence. Contextual evidence that is not tied to an artwork is counted by distinct records.

\textbf{Selecting the entry point and anchor artwork.} The planner considers recurring content under the four lenses, MyMuEx triggers, and contextual evidence when selecting an entry point. If none meets the recurrence threshold, the report begins with a single artwork. Among the four lenses, the leading lens is selected by the number of distinct supporting artworks, with utterance count breaking ties and a fixed CCO chapter ordering resolving any remaining tie. Among works supporting that lens, the planner selects the anchor artwork with the most lens-specific evidence, using broader activity around the work only as a tie-break. These rules determine where the report begins from the evidence available in the current record. Visitors can switch lenses to examine different relationships within the same collection.

\textbf{Composing the report.} An unchanged underlying record produces the same narrative plan. The report establishes the accumulated collection, presents recurrent evidence, introduces an anchor work, and shows relationships under different lenses. Each step appears only when supported by the available evidence. The resulting sequence can therefore vary across visitors while remaining stable across repeated runs over the same records.

\textbf{Organizing works through the four lenses.} Each lens organizes the selected artworks using different attributes. The \LensSubject{} view connects works through shared controlled catalogue subject terms. Visitor utterances and CCO annotations determine which works enter the narrative, while catalogue metadata determines the connections between them. A connection in this view therefore indicates a catalogue relationship between discussed works and does not imply that the visitor explicitly stated that relationship. \LensPhysical{} organizes works around properties such as color, material, or technique, with color and lightness measured from the visitor's photographs. \LensStylistic{} arranges works by time, movement, or period, and \LensCreator{} groups works by artist. Visitors can use these views to examine different relationships among the works included in the report.{\emergencystretch=6em\par}

\textbf{Generating and checking report text.} The language model receives selected facts after the planner has fixed the lens, anchor work, grouping, and supporting evidence. Its role is limited to phrasing supported observations. Generated text is checked for unsupported claims, speculative language, and statements that characterize stable visitor traits. If the text fails these checks, MyMuse uses deterministic fallback wording, retaining the planned content without relying on the rejected text.

\subsection{Implementation}
\label{sec:implementation}

MyMuse is implemented as a web application for desktop browsers. The Personal Museum is rendered with CSS 3D transforms, and visitors walk through the room with the arrow keys and look around with the mouse. A FastAPI server with a SQLite database serves the interface and stores catalogue records, visitor and system turns, annotations, and interaction logs. MyMuse's own language-model calls use \texttt{gpt-5.4-mini} through the OpenAI API, and classification and annotation calls request temperature~0. The ArtRAG component keeps \texttt{gpt-4o-mini} for its internal retrieval steps, including image-based reranking. The response model receives text only. The visitor's photograph is used by ArtRAG's image-based reranking, by CLIP retrieval, and for color measurement in the report. CLIP ViT-B/32 embeddings \cite{radford2021} are computed locally for the photographed artworks and the related-artwork corpus.

For each visitor turn, MyMuse determines whether the message asks for an answer, related artworks, or both, and which retrieval dimensions it names. It then assembles the response context (Section~\ref{sec:grounded-inquiry}), generates the reply, and stores the utterance and the reply as separate records together with the sources used for the reply.

Photographs are linked to catalogue records before they are placed in the Personal Museum. MyMuse does not identify artworks from photographs. Each record carries its source and a confidence level, both of which are included in the response context. Appendix~\ref{app:system} lists the inputs of each function and whether it requires a catalogue match.

In the study (Section~\ref{sec:user-study}), participants used MyMuse through this web application, with the server running on a single machine (Apple M4 Max, 128~GB of memory). Across the MyMuse sessions, it returned 332 responses, with a median server-side response time of 3.3~s (95th percentile: 8.7~s).

\section{Technical Evaluation}
\label{sec:tech-eval}

We evaluated artwork-level answer generation on Artpedia, agreement between human and model annotations at the lens level, and report-composition stability across repeated evaluations of unchanged records.

\subsection{Artwork-Level Answer Generation}
\label{subsec:eval-artpedia}

To assess the answer generation that supports artwork-level inquiry, we compared MyMuse and ArtRAG~\cite{wang25} using \texttt{gpt-4o-mini} on a fixed subset of 310 Artpedia~\cite{stefanini19} test items with recoverable images. Results varied across metrics: for example, MyMuse obtained higher ROUGE-L~\cite{lin04} ($27.49$ vs.\ $22.74$) but lower CLIPScore~\cite{hessel21} ($77.49$ vs.\ $84.88$) than ArtRAG, with both metrics multiplied by 100. With \texttt{gpt-5.4-mini} as the generation model for both systems, as in the deployed system (Section~\ref{sec:implementation}), the same pattern held (ROUGE-L $23.71$ vs.\ $20.28$; CLIPScore $77.36$ vs.\ $83.23$). The configurations differed in their inputs and prompting. Full methods, results, response lengths, and analyses by ArtRAG knowledge-graph coverage are reported in Appendix~\ref{app:artpedia}.

\subsection{Annotation Agreement}
\label{subsec:eval-annotation}

We evaluated agreement on the four lenses used in artwork-level inquiry and collection-level organization. Two researchers independently annotated all 332 visitor utterances from the MyMuse condition of Study~1 using a shared codebook, without access to the model annotations. We mapped each coder's and the model's CCO element assignments to four binary presence/absence labels for each utterance. Elements outside the four lenses did not contribute to these labels. For comparison with the model, we combined the human annotations by treating a lens as present when either coder assigned it.

Inter-coder agreement was pooled Cohen's $\kappa = .85$ ($94.8\%$ agreement), with per-lens values ranging from $.77$ to $.89$. Agreement between the model and the combined human labels was pooled $\kappa = .70$ ($89.4\%$ agreement), with per-lens values ranging from $.65$ to $.77$. The model matched the complete set of human-assigned lenses for $64.5\%$ of utterances.

Model--human agreement remained $\kappa = .70$ after excluding the 20 utterances flagged as uncertain by either coder. In a separate analysis, excluding utterances for which both the model and the human reference assigned no lens yielded $\kappa = .67$ across 260 utterances. These results indicate agreement on many individual lens labels, with differences in complete lens sets that could affect which records are grouped under each lens. Detailed lens-level results and sensitivity analyses are reported in Appendix~\ref{app:annotation-agreement}.

\subsection{Report Composition}
\label{subsec:eval-setup}

We examined whether report composition remained consistent across repeated evaluations of the same records and whether post-generation verification reduced variability. We compared three conditions using the records of the sixteen participants in the user study (Section~\ref{sec:user-study}).

In \textbf{A}, a language model selected the report composition under prompt constraints. \textbf{B} applied MyMuse's verifier to the corresponding report generated in A. In \textbf{C}, MyMuse's deterministic planner selected the composition, and the language model phrased the planned content. The A--B comparison assessed changes after verification, while B--C compared model-selected and deterministically planned compositions.

All conditions received the same participant-level records. A and B used the same output schema and applicable narrative constraints. We evaluated 16 participants $\times$ 3 conditions $\times$ 10 repetitions $\times$ 2 temperature settings, yielding 960 condition evaluations. Language generation used \texttt{gpt-5.4-mini}. For C, we bypassed the production cache to evaluate repeated runs. Prompts and execution details are provided in Appendix~\ref{app:report-composition}.

\begin{table}[t]
  \caption{Stability across ten repetitions on identical records at the default temperature. Cells report means over sixteen participants (sample SD). Jaccard values are averaged over pairs of retained repetitions within each participant. Pattern normalization combines CCO chapter mappings, MyMuEx trigger mappings, and normalization of unmatched phrases, as detailed in Appendix~\ref{app:report-stability-methods}.}
  \label{tab:eval-stability}
  \small
  \begin{tabular}{lccc}
    \toprule
    \textbf{Measure} & \textbf{A} & \textbf{B} & \textbf{C} \\
    \midrule
    Entry-lens disagreement     & 0.246 (0.214) & 0.240 (0.213) & 0.000 \\
    Anchor-artwork disagreement & 0.304 (0.207) & 0.304 (0.207) & 0.000 \\
    Artwork-set Jaccard         & 0.888 (0.145) & 0.888 (0.145) & 1.000 \\
    \rowcolor{ourrow}
    Pattern-set Jaccard         & 0.255 (0.166) & 0.254 (0.163) & 1.000 \\
    \bottomrule
  \end{tabular}
\end{table}

In A and B, selected artwork sets overlapped more consistently across repetitions than normalized pattern sets (Table~\ref{tab:eval-stability}). Artwork-set Jaccard was $0.888$ in both conditions, while pattern-set Jaccard was $0.255$ in A and $0.254$ in B. Post-generation verification produced little change in these measures under the tested conditions. Lowering the temperature also left variability in model-selected pattern sets. Results from runs in which \texttt{temperature=0} was requested are reported in Appendix~\ref{app:eval-temp}.

In C, all measured composition decisions were identical across repetitions, as expected from deterministic planning. With language generation disabled, the fallback described in Section~\ref{sec:report} produced structurally complete reports for all sixteen participant records without model calls during report generation.

\subsection{Scope}
\label{subsec:eval-scope}

The annotation results concern assignments to the four lenses. This analysis does not assess individual CCO elements, MyMuEx triggers, or relations between annotations. Artpedia scores measure correspondence with reference descriptions and images, without directly verifying whether retrieved sources support each generated claim. The benchmark also contains no visitor-authored inquiry histories with which to assess personalization or reasoning across personal records. The report analysis measures composition stability across repeated evaluations of unchanged records. Section~\ref{sec:results-study2} examines recurring-interest awareness, cross-artwork connections, unexpected discovery, and perceived fidelity with participants.
\section{User Study}
\label{sec:user-study}

We evaluated MyMuse at the two levels of post-visit engagement it was designed to support: inquiry around individual artworks and reflection across an accumulated collection. At the individual-work level, we examined whether carrying forward visitors' own questions and reflections could support more personally grounded revisiting and further inquiry than an informationally comparable per-artwork baseline. At the collection level, we examined whether reorganizing the same accumulated records across artworks could make recurring interests and cross-work relationships more legible while remaining grounded in what participants had recorded. We examined these questions through two within-subject studies conducted within a single session. Because the evaluation was conducted in a single session, our claims concern immediate effects rather than longer-term changes in interpretation or museum-going practice.

We chose this design to examine the point at which visitors return to photographs from earlier visits, where formative participants had difficulty recalling why they had photographed works (F1). Participants therefore used their own photographs from past visits rather than photographs from a visit arranged for the study.

\subsection{Study Setup}
\label{sec:study-setup}
\label{sec:study-participants}
\label{sec:study-conditions}
\label{sec:study-procedure}
\label{sec:study-measures}
\label{sec:study-analysis}

\paragraph{Participants and procedure}
We recruited 16 participants (P01--P16; 12 women, 4 men), aged 19--55 ($M = 26.5$, $SD = 8.3$). Most were undergraduate or graduate students, alongside two working professionals and one homemaker. Twelve reported no art-related academic or professional background, while four reported art-related academic or professional experience. All participants had visited a museum or gallery at least twice in the previous 12 months (2--3 visits: 5; 4--6: 4; 7 or more: 7). Self-rated art knowledge ranged from 1 to 7 ($M = 3.8$, $SD = 1.5$). Eligibility required participants to have at least 20 personally taken artwork photographs from prior museum or gallery visits. Participants from the formative study were not eligible. The study was approved by our institution's IRB. All participants provided informed consent and received \$20.

Each participant submitted 16 of their own artwork photographs from earlier visits. Photographs with recorded capture dates had been taken a median of 128 days (about 4.2 months) before the session. We randomly divided each participant's photographs into two non-overlapping sets of eight, one for each Study~1 condition. Using separate sets reduced carryover from revisiting the same artwork in both conditions. Researchers matched each photograph to its catalogue record using the information participants provided, photographed wall labels, and the museums' published records, and recorded each record's source and confidence level, so that both conditions had access to the same verified metadata and contextual information. Each session lasted approximately 90 minutes. Participants first completed an opening interview, then experienced both Study~1 conditions, directly compared the two systems, examined both Study~2 reports, and completed a closing interview. The approximately 10-minute opening interview covered participants' existing museum-going, information-seeking, and record-keeping practices, including what they attended to when viewing artworks and what they chose to photograph. This semi-structured interview provided context for interpreting their subsequent evaluations.

\paragraph{Conditions and tasks}
During the study, the conditions were referred to neutrally as ``System A'' and ``System B,'' and the reports as ``Report A'' and ``Report B.'' We independently counterbalanced the order of the two Study~1 conditions and the two Study~2 conditions, yielding four order combinations with four participants each.

\emph{Study 1: MyMuse interaction and baseline.}
Both conditions allowed participants to browse their submitted artworks, inspect catalogue information, ask free-form questions, leave personal reflections, and receive related-work recommendations. The baseline provided these capabilities through a conventional per-artwork chat interface. It retained participants' entries as per-artwork conversation history, and prior entries did not inform subsequent responses. The MyMuse interaction provided the interface described in Section~\ref{sec:system}. It reused accumulated records to contextualize later responses and spatially organized participants' artworks and records within a personal museum. Thus, the conditions shared the same core informational and exploratory capabilities but differed in how participants' personal records were organized and reused.

Participants used each condition for approximately 15 minutes in their counterbalanced order, using the assigned set of eight photographs. Before each condition, the researcher briefly introduced the available interactions. Participants then explored their artworks freely by browsing the collection, asking questions, viewing related works, and leaving reflections as they wished. We did not prescribe specific questions or exploration targets because both conditions were intended to support self-directed post-visit exploration rather than completion of a predefined task. Immediately after each condition, participants completed a short semi-structured interview followed by the corresponding questionnaire. The interview focused on their overall experience, how they had used the system, whether it supported revisiting or further exploring prior encounters, and how they might use such a system after a museum visit. After evaluating both systems individually, participants completed a comparative semi-structured interview followed by a direct-comparison questionnaire. This interview focused on perceived differences between the systems, their useful aspects, and participants' overall preferences.

\emph{Study 2: Structured Record and Narrative Report.}
Both conditions were derived from the records participants had produced during the MyMuse interaction. The \emph{Structured Record} presented the accumulated record per artwork, grouping each photograph and its catalogue information with the participant's associated questions, reflections, and exploration records. The \emph{Narrative Report}, described in Section~\ref{sec:report}, selected and organized evidence from these records into an interactive cross-work narrative, with access to supporting visitor-authored entries and further artwork exploration. The two reports therefore shared a source record but differed in content organization and interaction. Study~2 examined participants' experiences of these two report designs as a whole.

Participants examined the two reports in their counterbalanced order, spending approximately 10 minutes on each report. For the Narrative Report, MyMuse automatically selected one of the four interpretive lenses from the participant's accumulated records as an entry point (Table~\ref{tab:lens-assignment}). Participants first examined the report through this lens and were then free to explore the remaining lenses and other parts of the report. For the Structured Record, the researcher briefly introduced its per-artwork organization, after which participants explored the report freely. After each report, participants completed the same four-item questionnaire. The session concluded with a comparative interview about the two reports and the overall experience. We asked what participants noticed about their own interests and relationships across artworks, what they found new or unexpected, which report they preferred and why, what each format would be useful for, and what they would want the system to support beyond the initial encounter.

\begin{table}[t]
\centering
\small
\caption{Interpretive lens selected as each participant's entry point into the Narrative Report.}
\Description{A table showing the Narrative Report entry-lens assignments for sixteen participants. Creator is selected for P10 and P16, totaling two participants. Physical Characteristics is selected for P01, P05, P07, P11, and P14, totaling five. Stylistic and Chronological is selected for P02 and P09, totaling two. Subject is selected for P03, P04, P06, P08, P12, P13, and P15, totaling seven. Each participant has one selected entry lens.}
\label{tab:lens-assignment}
\begin{tabular}{l l c}
\toprule
\textbf{Lens} & \textbf{Participants} & $n$ \\
\midrule
\lenstag{Creator}{Creator} & P10, P16 & 2 \\
\lenstag{Physical}{Physical} & P01, P05, P07, P11, P14 & 5 \\
\lenstag{Stylistic}{Stylistic} & P02, P09 & 2 \\
\lenstag{Subject}{Subject} & P03, P04, P06, P08, P12, P13, P15 & 7 \\
\bottomrule
\end{tabular}
\end{table}

\begin{figure*}[t]
    \centering
    \includegraphics[width=\textwidth]{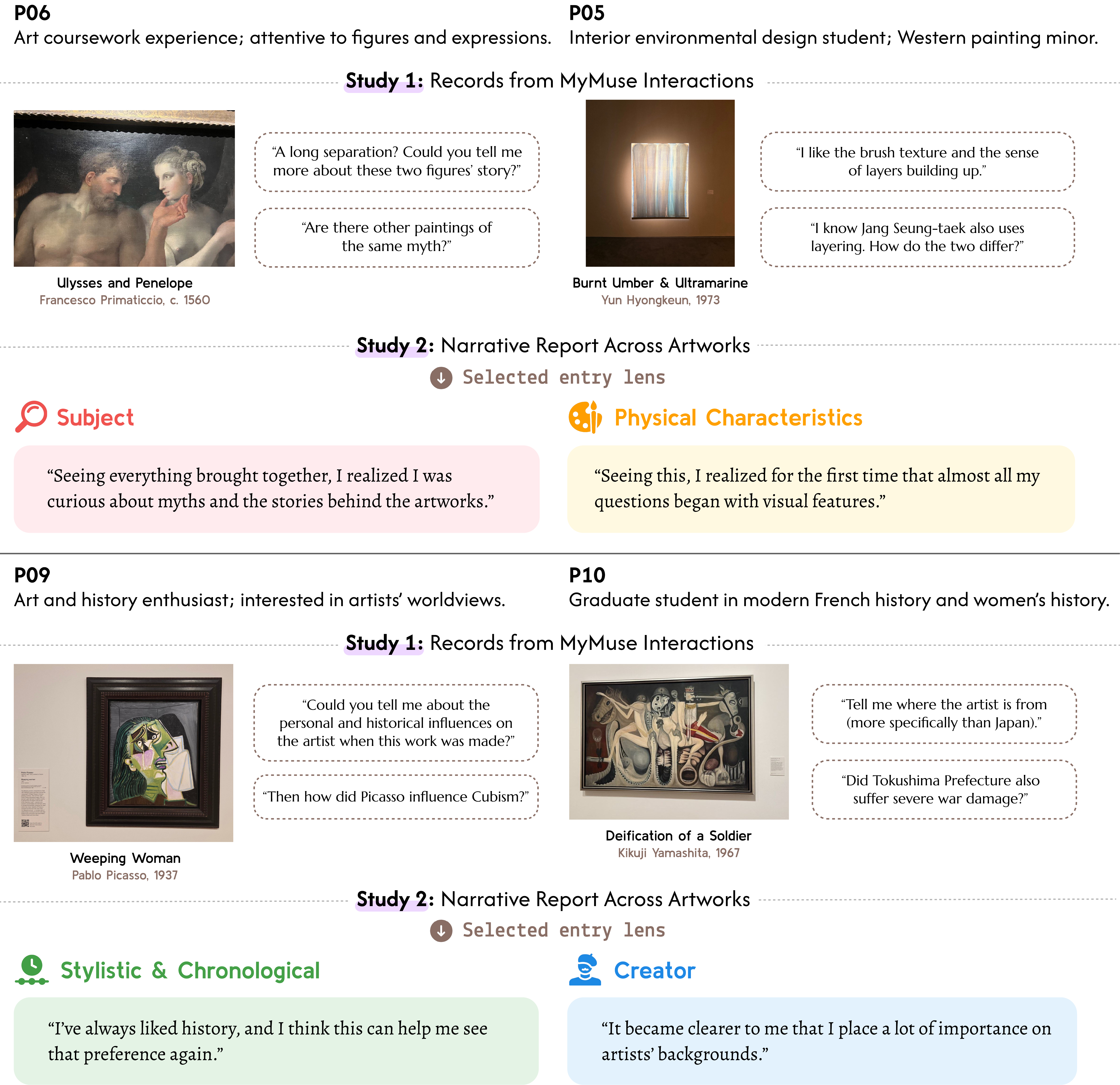}
    \caption{Illustrative participant cases across the two study components. The upper part of each case shows examples of records participants produced while revisiting individual artworks in the MyMuse condition of Study~1. The lower part of each case shows the entry lens selected from each participant's accumulated records in the Study~2 Narrative Report and an example of what the participant reported noticing when examining records across artworks. Participant descriptions provide contextual background and were not used to determine the selected lens. Appendix~\ref{app:example-exchange} shows P09's exchange with MyMuse's responses.}
    \Description{Four participant cases arranged in a two-by-two grid, with P06 and P05 above P09 and P10. Each case includes participant background, an artwork photograph, and two Study 1 utterances, followed by the Study 2 entry lens selected from accumulated records and an interview reflection. P06 discusses mythological stories in Ulysses and Penelope, and their entry lens is Subject. P05 discusses texture and layering in Burnt Umber and Ultramarine, and their lens is Physical Characteristics. P09 asks about historical influences on Weeping Woman, and their lens is Stylistic and Chronological. P10 asks about the artist's origins and wartime context of Deification of a Soldier, and their lens is Creator. The interview excerpts describe participants recognizing or clarifying their interests across artworks.}
    \label{fig:study-cases}
\end{figure*}

\paragraph{Measures}
Unless otherwise noted, questionnaire items used 7-point Likert scales (1 = strongly disagree, 4 = neutral, 7 = strongly agree).

\emph{Study 1.}
Six shared items captured complementary aspects of participants' experience with the two systems. \emph{Reflective Outcomes} comprised: (1)~\emph{re-experiencing}, whether using the system vividly brought the prior museum visit to mind; (2)~\emph{elaboration}, whether it helped participants develop their thoughts about an artwork more concretely; and (3)~\emph{connections among responses}, whether it helped them notice commonalities or connections across their responses to different works. \emph{System Quality and Intention} comprised: (4)~\emph{answer relevance}, whether system responses matched what participants wanted to know; (5)~\emph{engagement}, whether they wanted to further explore their artworks and records; and (6)~\emph{intention to use}, whether they would use the system again after a museum visit.

A seventh shared item measured \emph{findability}, or whether participants could easily find and inspect the artworks and records they wanted. We report it separately as a usability check because it concerns interface operation rather than the outcomes above. The MyMuse questionnaire additionally contained diagnostic items assessing personalized Q\&A informed by previous records and spatial reconstruction. Because these features had no direct counterparts in the baseline, we analyze these items descriptively. After using both systems, participants completed three bipolar 7-point direct-comparison items. Two compared the systems on \emph{connections among responses} and \emph{re-experiencing}, with written explanations for each judgment, and the third asked which system they would prefer to use after a museum visit. Scale endpoints remained fixed regardless of presentation order.

\emph{Study 2.}
Three items captured collection-level reflective outcomes: (1)~\emph{recurring-interest awareness}, whether the report helped participants notice interests recurring across multiple works; (2)~\emph{cross-artwork connection awareness}, whether it helped them recognize connections among their experiences or reactions to different artworks; and (3)~\emph{unexpected discovery}, whether it revealed interests or relationships they had not previously considered. A fourth item assessed \emph{perceived fidelity}, whether the presented content accurately reflected the records participants had actually left. We report it separately as a grounding check rather than as a measure of collection-level reflection. Although Study~1 also included a connection-related item, the two measures targeted different levels: Study~1 concerned connections noticed while revisiting individual works, whereas Study~2 concerned connections surfaced across the accumulated collection. We therefore treat them as related but distinct outcomes.

\paragraph{Data analysis}
For all Wilcoxon analyses, we used two-sided signed-rank tests, excluded zero differences, and used the normal approximation in the presence of ties. We report descriptive statistics and effect sizes as $r = |Z|/\sqrt{N}$, where $N$ is the number of participants. For Study~1, the six shared outcome items were analyzed individually. We report unadjusted $p$-values with effect sizes and interpret these item-level comparisons as indicative rather than confirmatory. Holm-adjusted $p$-values are provided in Appendix~\ref{app:user-study-statistics}. Findability was treated separately as a usability check, and the reported MyMuse-specific diagnostic items were summarized descriptively because no corresponding baseline measures existed. For Study~2, the three collection-level reflective outcomes were analyzed individually, with Holm correction applied across the three comparisons. Perceived fidelity was examined separately as a grounding check.

The three direct-comparison items were single bipolar ratings rather than paired condition-specific ratings. We therefore tested each against the scale midpoint using the same Wilcoxon procedure and report directional counts and effect sizes. Unadjusted $p$-values are reported in the main text, with Holm-adjusted values in Appendix~\ref{app:user-study-statistics}.

Interviews were audio-recorded with participant consent and transcribed. We analyzed the transcripts and written explanations using reflexive thematic analysis informed by the study questions. As in the formative study, the analysis was interpretive rather than a classification task, so we do not report inter-rater agreement. The qualitative analysis was used to interpret the measured outcomes and contextualize cases that diverged from aggregate ratings.

Figure~\ref{fig:study-cases} provides four illustrative cases showing how records produced during the MyMuse interaction in Study~1 were later reorganized and revisited through the Narrative Report in Study~2.

\subsection{Study 1: Returning to Individual Works}
\label{sec:results-study1}

\begin{figure}[t]
    \centering
    \includegraphics[width=\linewidth]{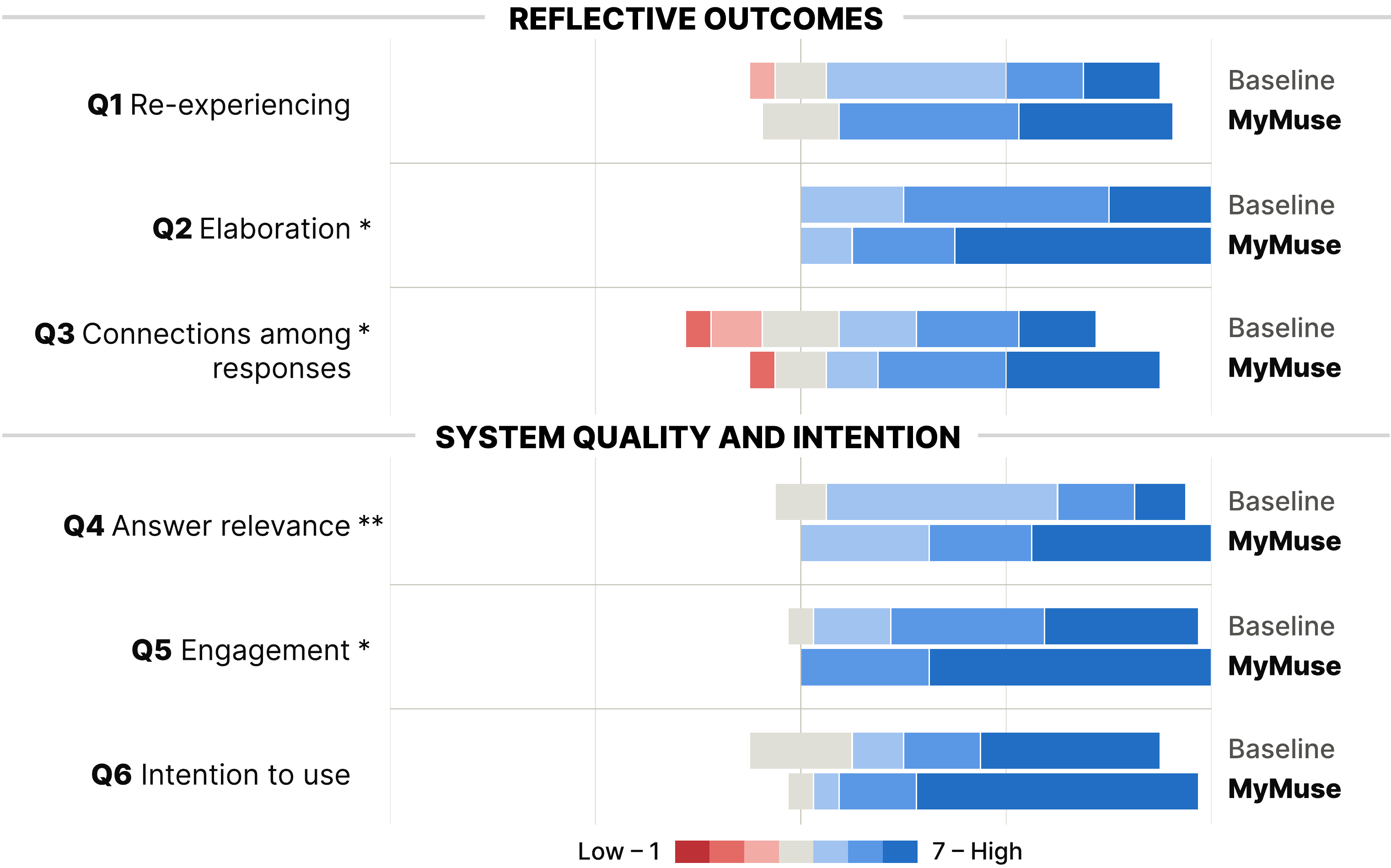}
    \caption{
    Study~1 ratings for the six shared outcomes under the baseline and MyMuse conditions.
    Diverging bars show the distribution of responses on 7-point Likert scales
    (1 = strongly disagree, 7 = strongly agree), grouped into
    \emph{Reflective Outcomes} and \emph{System Quality and Intention}.
    Asterisks indicate unadjusted two-sided Wilcoxon signed-rank tests
    ($^{*}p < .05$, $^{**}p < .01$).
    These item-level tests are exploratory, as described in
    Section~\ref{sec:study-analysis}. Holm-adjusted values are reported in the Appendix.
    }
    \Description{
    Diverging Likert charts compare baseline and MyMuse ratings on six Study~1
    outcomes: re-experiencing, elaboration, connections among responses,
    answer relevance, engagement, and intention to use.
    Ratings generally shift toward higher values for MyMuse.
    }
    \label{fig:study1}
\end{figure}

Figure~\ref{fig:study1} summarizes the six shared outcomes, and Appendix~\ref{app:user-study-statistics} reports the full statistics. As described in Section~\ref{sec:study-analysis}, we interpret the unadjusted item-level and direct-comparison tests as exploratory.

\subsubsection{Participants valued responses that addressed their observations and questions}

Participants rated answer relevance higher for MyMuse than for the baseline ($M = 6.12$ vs.\ $5.31$; $p = .006$, $r = .69$), the only shared outcome that remained significant after Holm correction, and rated the MyMuse-specific personalized Q\&A item positively ($M = 6.06$). They described relevant responses as addressing what they wanted to understand about a work. After P13 commented on the figure's gaze in Rombouts's \emph{Lute Player}, the response explained that the figure was tuning the instrument. P13 said, \textit{``It gave me the answers I really wanted and explained the reasons for my impressions.''} Others reported little difference between conditions (P04, P08) or found the answers useful without perceiving them as personally tailored (P15), and P10 described an answer that concentrated on style despite their questions about context. Because participants also contributed differently across conditions, with P13, for example, entering more reflections in MyMuse, we cannot attribute perceived relevance specifically to the reuse of prior records.

\subsubsection{Participants associated spatial browsing with vivid revisitation}

Re-experiencing ratings did not differ significantly between conditions ($M = 6.00$ vs.\ $5.31$; $p = .065$), although 13 of 16 participants favored MyMuse in the direct comparison ($p = .001$, $r = .81$), and the MyMuse-specific item on spatial organization was rated positively ($M = 6.06$). Several participants associated vividness with moving through a room of photographs they had selected themselves. P15 explained, \textit{``I could move back and forth in that space. I think that made the memories come back more vividly.''} Revisiting also involved learning about works whose earlier impressions were hard to recall. P10 said that neither system brought back the emotions they had felt when first seeing a work. Others found the spatial presentation less helpful. P03 found revisiting less vivid because the room differed from the museum they had visited, and P05 preferred a familiar photograph layout with visible contextual information. P13 disliked moving through the room but preferred MyMuse overall, attributing better recollection to its larger artwork display. A preference for MyMuse therefore did not necessarily indicate a preference for spatial navigation.

\subsubsection{Participants articulated and developed their responses to artworks}

Elaboration was rated highly in both conditions and higher for MyMuse ($M = 6.50$ vs.\ $6.00$; $p = .046$, $r = .50$). The item concerns perceived support rather than the amount or type of text entered. Participants related what they wrote to how the artwork was presented. P05 wrote more immediate reactions in MyMuse because \textit{``it felt like viewing art in a museum''} and asked more questions in the baseline. P01 similarly entered more memories and earlier impressions in MyMuse, and P13 attributed their focus on present feelings to the larger artwork and smaller question window.

\subsubsection{Viewing works together prompted some cross-work comparisons}

Participants rated support for recognizing connections among their responses higher for MyMuse ($M = 5.75$ vs.\ $5.00$; $p = .032$, $r = .54$), and 12 of 16 favored it in the direct comparison ($p = .002$, $r = .76$), although this remained its lowest-rated shared outcome. The comparisons participants described mostly concerned visible differences and exhibition contexts. P16 noticed color differences among three fan paintings only when viewing them together, and P12 and P11 found grouping by museum or exhibition helpful for following connections. Participants also reconsidered their photographic choices, as when P04 noticed a tendency to photograph famous works, but did not consistently identify how their questions and reflections related across the collection. P08 wanted more help noticing such connections from either system.

\subsubsection{System preferences reflected participants' intended uses}

Engagement was rated higher for MyMuse ($M = 6.69$ vs.\ $6.06$; $p = .026$, $r = .56$), whereas intention to use ($p = .093$) and findability ($p = .145$) did not differ significantly. Asked which system they would use after a museum visit, 12 participants chose MyMuse and four chose the baseline ($p = .049$, $r = .49$). Three of these four had favored MyMuse on at least one of the preceding direct comparisons. They weighed anticipated archiving and retrieval needs: P04 wanted a simple way to organize a large camera roll, P14 preferred the baseline for keeping an ongoing record of impressions, and P10 explained, \textit{``For archiving and finding things again later, [the baseline] is actually more convenient.''}

% ---------------------------------------------------------------------

\subsection{Study 2: Reading the Accumulated Collection}
\label{sec:results-study2}

\begin{figure}[t]
    \centering
    \includegraphics[width=\linewidth]{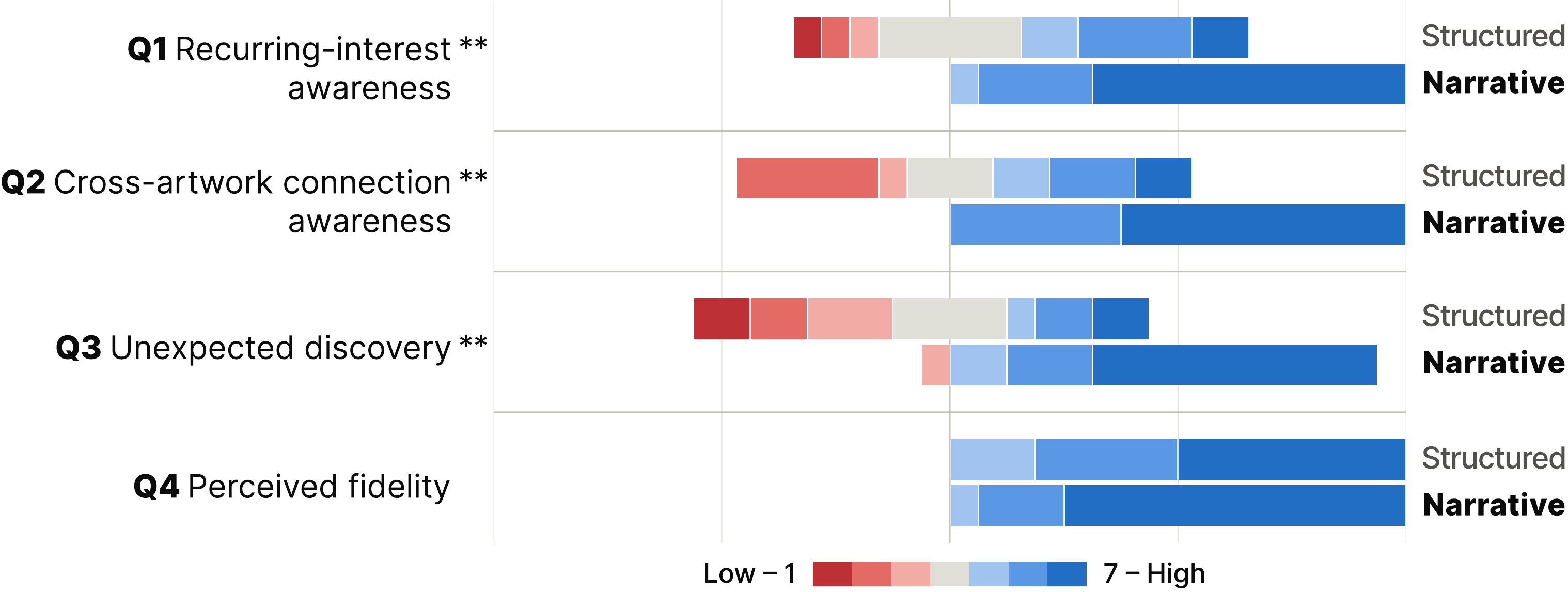}
    \caption{
    Study~2 ratings for the Structured Record and Narrative Report.
    Diverging bars show the distribution of responses on 7-point Likert scales
    (1 = strongly disagree, 7 = strongly agree).
    For the three collection-level reflective outcomes, asterisks indicate
    significance based on Holm-adjusted $p$-values from two-sided Wilcoxon
    signed-rank tests
    ($^{*}p < .05$, $^{**}p < .01$).
    Perceived fidelity is shown separately as a grounding check and was not
    included in the Holm correction.
    }
    \Description{
    Diverging Likert charts compare the Structured Record and Narrative Report
    on recurring-interest awareness, cross-artwork connection awareness,
    unexpected discovery, and perceived fidelity.
    Ratings for the three reflective outcomes are substantially higher for the
    Narrative Report, while perceived fidelity is high in both conditions.
    }
    \label{fig:study2}
\end{figure}

Figure~\ref{fig:study2} summarizes the four Study~2 measures, and Appendix~\ref{app:user-study-statistics} reports the full statistics. The three collection-level reflective outcomes were tested with Holm correction, and perceived fidelity was assessed separately.

\subsubsection{Participants recognized recurring interests and connections across works}

The Narrative Report was rated higher than the Structured Record for cross-artwork connection awareness ($M = 6.62$ vs.\ $4.19$; $p_{\mathrm{Holm}} = .003$, $r = .83$) and recurring-interest awareness ($M = 6.62$ vs.\ $4.62$; $p_{\mathrm{Holm}} = .005$, $r = .76$). Participants recognized relationships among both the works and their own questions. P06 explained, \textit{``I thought there were almost no keywords that could bring these works together, but grouping them as women or girls made it easier to discover new commonalities,''} and recognized an attention to depicted stories and mythology. P16 found that questions left about individual works were more connected than expected, and P02 could see their questions about period and style at a glance. The Structured Record also prompted comparisons: ordering works by year led P10 to new questions about decolonization and war, and P05 newly recognized their attention to composition.

\subsubsection{Some participants reconsidered how they engaged with artworks}

Unexpected discovery was also rated higher for the Narrative Report ($M = 6.31$ vs.\ $3.88$; $p_{\mathrm{Holm}} = .005$, $r = .72$). Some participants reconsidered their viewing practices. P05 realized, \textit{``Almost all my questions started from visual aspects,''} and P11 recognized that color tended to attract their attention first. P15 realized that after visits they wanted to know more about the work and artist than about its visual qualities, a distinction they had begun to articulate earlier in the interview. Others recognized familiar interests without new discovery (P01, P09, P13), and P03 questioned whether their limited record supported the analysis. Participants also differed in whether they wanted an account of themselves: P08 compared the report to a personality-test result, whereas P13 said, \textit{``When I look at paintings, I don't really need data analyzing me.''}

\subsubsection{Participants rated fidelity highly while questioning some derived relationships}

Perceived fidelity was high for both reports and did not differ significantly ($M = 6.69$ vs.\ $6.31$; $p = .083$). Participants recognized their own contributions. P06 said, \textit{``I was impressed that it actually excerpted and organized the parts where I had shared my impressions.''} They nevertheless questioned some of the classifications and comparisons presented around those contributions. P05 and P11 questioned comparisons based on color distribution, and P05 suggested materials and techniques as more useful grounds, whereas P04 valued the color-focused displays. P01 noted that an existing interest in war was absent because they had not discussed it with the system.

\subsubsection{Participants valued the Structured Record for review and reuse}

Some participants preferred the Structured Record for how they expected to use their records. P05 and P14 anticipated drawing on it when writing reports or exhibition reviews, and P06 expected to check earlier questions and impressions quickly. Preferences did not always follow reflective ratings. P03 rated the Narrative Report at least as highly on all three reflective outcomes and P13 favored it on average, yet both preferred the Structured Record, citing limited material and little need for analysis. P10 found the Narrative Report more helpful for recognizing interests across works but preferred the Structured Record as easier to revisit as records accumulate, and P12 valued the Narrative Report's visual links but found it harder to recall earlier questions through it. Participants could thus value the Narrative Report for reflection and still prefer the Structured Record for review.
\section{Discussion}
\label{sec:discussion}

\subsection{Inquiry and Reinterpretation Around Individual Artworks}
\label{subsec:discussion-scales}

Participants' accounts suggest that contextual information was useful when they could relate it to observations they had already made. P06's conversation about \emph{Ulysses and Penelope} illustrates this process. After describing the figures' seemingly misaligned gazes, P06 asked for more detail about the separation and reunion introduced in the response. They then elaborated on the poignancy of the figures' expressions and requested other depictions of the same myth. These follow-up questions showed how P06 used the contextual information to develop an earlier observation and pursue comparisons with other works.

Recollection and new interpretation also contributed differently to participants' experiences. P10 described learning more about artworks while remaining unable to recall their original emotions. This account suggests that returning to a photograph may be valuable as an opportunity to reconsider a work even when recollection remains incomplete. Supporting revisitation therefore involves attending to what visitors want to understand in the present as well as what they hope to remember from the original encounter.

\subsection{Interpreting Interests Across Personal Records}
\label{subsec:discussion-study2-interests}

The two studies suggest a distinction between comparing artworks and examining relationships among the records associated with them. In Study~1, P16 noticed differences among fan paintings viewed together, while P08 still found it difficult to recognize connections among their responses. In Study~2, the Narrative Report received higher ratings on the three collection-level reflective outcomes than the Structured Record. The Structured Record also prompted participants to recognize interests and relationships across works. The Narrative Report combined evidence selection, cross-work grouping, explanations, visual links, and access to supporting records. The higher ratings concern participants' experiences of this combined presentation and do not isolate the contribution of its individual features.

Participants' accounts indicate that this reflection included clarifying existing interests as well as recognizing previously unnoticed connections. P06 recognized an interest in stories and mythology across works they had initially regarded as having little in common. P11 already knew that visual qualities mattered to them, but the report helped them identify color as something that often attracted their attention first. For P15, the report clarified a distinction between visual enjoyment during a visit and information seeking afterward that they had already begun to articulate. These experiences suggest that reading across the questions and interpretations recorded during inquiry can help visitors reflect on what they responded to and describe their interests more specifically, even when the resulting interpretation is not entirely new.

The report's descriptions remained dependent on what participants had chosen to record. P01 explained that an existing interest in war was absent because they had not discussed it with the system. Perceived fidelity assessed how well the presentations reflected recorded contributions, so positive ratings on this measure did not establish that the reports represented participants' interests comprehensively. The formative finding that photographs could leave their capture reasons unclear further emphasizes the importance of what visitors articulate around an image. These observations support interpreting report patterns in relation to the selected works and exchanges. Access to the underlying contributions could help visitors assess how far a presented relationship reflects what they intended to express.

Participants also differed in which comparisons they found informative. P04 valued the color-focused displays, while P05 and P11 questioned the usefulness of some color comparisons. P05 proposed materials and techniques as more relevant grounds for relating works. These accounts suggest that a comparison can be meaningful to one visitor and less useful to another because they attend to different aspects of artworks. Offering alternative grounds for comparison may help visitors examine the collection in ways that correspond to their current questions \cite{thudt12,windhager19,yee03}.

\subsection{Supporting Revisitation and Review}
\label{subsec:discussion-purposes}
\label{subsec:discussion-study2-review}

Participants valued the interfaces for different activities. In Study~1, several associated MyMuse's spatial presentation with immersion, movement, and viewing artworks together, while others preferred the baseline's simpler presentation for organizing photographs and finding them again. P03 found revisitation less vivid because the room differed from the museum they had visited. This response suggests that expectations about the viewing environment can affect the value of spatial presentation. The conditions also differed in conversational support and information presentation, so these accounts describe responses to the interfaces as a whole and cannot establish the contribution of dimensionality alone.

Study~2 similarly showed that recognizing relationships and reviewing earlier contributions could lead to different preferences. P10 found the Narrative Report more helpful for recognizing interests across works, yet preferred the Structured Record because they expected a growing archive to become difficult to revisit through the narrative presentation. This preference concerned anticipated use beyond the immediate reading experience. It suggests that a presentation valued for reflection may still need additional support for locating and reviewing particular records.

P12's experience identifies a more specific difficulty with such review. They valued the Narrative Report's visual links and explanations but found it harder to recall their earlier questions through that presentation. Supporting records were available, yet their presence did not necessarily make earlier exchanges easy to review. Clearer access to full exchanges and a compact overview of contributions could help visitors return to a particular question while reading a report. Such access may be especially useful when the purpose of a later visit to the archive differs from the relationship emphasized in the narrative.

\subsection{Using Records Across Activities and Visits}
\label{subsec:discussion-future-engagement}
\label{subsec:discussion-study2-future-use}

Participants anticipated using their records as reference material for learning, creative practice, and writing. P05 described studying exhibition display through their records and drawing on that material when planning or participating in future exhibitions. P11 connected possible uses for recording interpretations and building knowledge to an existing practice of referring to photographed techniques in their own work, and also described a possible role in art criticism. P05 and P14 anticipated using the Structured Record when writing exhibition reviews, reports, or other texts. These accounts suggest that retaining personal responses alongside contextual information could support the development of an interpretation or argument in another activity. These uses were anticipated or drawn from existing practices. Their contribution to subsequent work remains to be evaluated.

Visitors may also want to record a thought without continuing a dialogue. P10 described wanting to preserve why they had photographed a work and finding an additional empathetic system response tiring. This feedback suggests separating the act of saving a contribution from the decision to request an answer. Options to record an impression without an immediate response, request context when needed, and choose the depth of an explanation could accommodate different purposes within the same interaction.

Participants anticipated different preferences for when recording and inquiry would fit into a visit. P09 and P15 preferred to protect their attention during viewing and return to questions or records afterward. P10 expected direct capture and access during a visit to reduce the effort of uploading photographs separately at home. A future system could accommodate these preferences by allowing a photograph or brief question captured on a phone to remain available for more sustained inquiry afterward. This would extend the present post-visit setting while leaving visitors to decide when recording and contextual support fit their viewing activity.

Accumulated records could also inform what visitors choose to encounter next. P08 imagined consulting reports from earlier exhibitions when choosing another exhibition, and P12 described considering future museum visits in relation to works they had liked. Together with requests for related artworks, these accounts suggest investigating how recorded questions and interests might inform subsequent visits. A future system could connect questions about a technique, subject, historical context, or curatorial choice to relevant exhibition information and explain which contributions informed each suggestion. Such suggestions could give visitors options for pursuing an existing interest or exploring another direction. We did not evaluate exhibition discovery or subsequent attendance, so the quality of these suggestions and their effects on later choices remain open questions.

\subsection{Implications for Systems Built on Personal Records}
\label{subsec:discussion-implications}

The two studies point to a relationship between inquiry and reflection that extends beyond museum visits. The questions and interpretations that people add while examining individual items are themselves evidence of what they attend to, so a system can treat inquiry as the source of later reflection rather than as a separate activity. MyMuse's reports could describe only what participants had articulated (Section~\ref{subsec:discussion-study2-interests}). Systems for personal photographs, learning journals, or design reference collections \cite{yin25,nuckles20,herring09} would likewise need ways for people to record why an item matters before collection-level summaries can reflect their interests.

When the same language model both answers questions and summarizes a person's records, the boundary between what the person said and what the system said is easily lost. MyMuse treated only visitor-authored text as evidence about the visitor and presented catalogue-based connections as catalogue relationships rather than as the visitor's statements. Participants rated the reports as faithful to their records while questioning some derived relationships (Section~\ref{sec:results-study2}). Systems that answer questions over personal data with language models \cite{li25} could similarly keep authored, retrieved, and generated content distinguishable, so that a summary of a person's interests does not absorb the system's own suggestions.

Collection-level claims about a person also call for more conservative composition than item-level answers. Counting recurrence across distinct items and fixing a report's composition before generating its wording kept the report structure identical across repeated runs, whereas model-selected compositions varied (Section~\ref{subsec:eval-setup}). Approaches that select facts for data narratives \cite{shi21,wang20} could adopt similar constraints when the data describe a person rather than a dataset.

\section{Limitations and Future Work}
\label{sec:limitations}

\textbf{Following records across repeated visits.} Our studies examined immediate effects in a single session with 16 participants, most of them students and all of them recent museum visitors who photograph artworks, and the Study~2 reports drew on records around eight artworks. A longitudinal deployment with more varied visitors would show whether people return to their records once the novelty of a first session has passed, how their interpretations change across visits, and whether the recurrences a report identifies persist as collections grow.

\textbf{Identifying artworks from visitors' own photographs.} Researchers prepared the catalogue records for the study (Section~\ref{sec:study-setup}), and MyMuse's artist- and date-dependent functions rely on them (Appendix~\ref{app:system}). An identification step that proposes candidate records from the photograph, a photographed wall label, and the time and place of the visit, and lets the visitor confirm or correct them, would let MyMuse work with newly taken photographs. Drawing on a wider range of institutional catalogues would also extend related-work retrieval beyond its predominantly Western corpus.

\textbf{Isolating what the narrative contributes.} Study~2 compared the Narrative Report as a whole, combining evidence selection, cross-work grouping, explanations, visual links, and guided sequencing, with a per-artwork record. Comparing it with a non-narrative cross-work view that presents the same evidence would show how much guided sequencing itself helps visitors recognize recurring interests.

\textbf{Making model-mediated steps inspectable.} Utterance annotation agreed with human coders at $\kappa = .70$ (Section~\ref{subsec:eval-annotation}), and responses combined catalogue records with retrieved context and model knowledge (Section~\ref{sec:grounded-inquiry}). Showing visitors the sources recorded for each response, evaluating factual accuracy on visitors' own questions, and examining how annotation differences change which records a report groups together would help visitors and designers decide when to rely on these steps.

\section{Conclusion}
\label{sec:conclusion}

We presented MyMuse, a post-visit personal museum that retains visitors' questions and interpretations alongside artwork photographs. Its design was informed by formative participants' difficulty recalling why they had photographed works and their interest in examining connections across their collections. MyMuse uses prior entries to inform subsequent responses and organizes accumulated records in a Narrative Report for comparison across works.{\emergencystretch=4em\par}

In a controlled within-subject evaluation with 16 participants, more participants favored MyMuse for vivid revisitation and cross-work connections in direct comparisons. The Narrative Report received higher ratings for recurring-interest awareness, cross-artwork connections, and unexpected discovery than the Structured Record. Participants described recognizing previously unnoticed connections and clarifying familiar interests. These findings suggest that organizing questions and interpretations across artworks can help visitors examine recurring interests expressed in their personal records.

\bibliographystyle{ACM-Reference-Format}
\bibliography{references}

\clearpage
% The appendix is set in one column: tables stay next to the text that
% discusses them, and the prompt listings keep their own line breaks.
\onecolumn
\raggedbottom
% One spacing around every appendix table.
\setlength{\intextsep}{12pt plus 2pt minus 2pt}
\setlength{\textfloatsep}{12pt plus 2pt minus 2pt}
\setlength{\floatsep}{12pt plus 2pt minus 2pt}
\appendix
\section{Additional Formative Study Materials}
\label{app:formative}

This appendix summarizes the formative interview topics, illustrates connections between initial codes and the four reported themes, and describes the separate CCO coverage analysis and MyMuEx trigger selection.

\subsection{Interview Topics}
\label{app:formative-interview}

The semi-structured interviews addressed museum-going, artwork photography, information seeking, recording, organization, and revisitation practices. Photo elicitation used ten photographs selected by each participant from their own museum or gallery visits. For each photograph, we discussed the reason for capture, what remained memorable, what information or aspects of the encounter were no longer recoverable, and the relevance of later knowledge or intended reuse.

Later in each session, we invited open-ended suggestions about how
a system for revisiting artwork photographs, recording reflections,
and asking questions could be organized. We distinguish these design
suggestions from accounts of participants' existing practices and
interests. Suggestions concerning spatial presentation informed the
design choices described in Section~\ref{sec:personal-museum}.

\subsection{Initial Codes and Theme Summary}
\label{app:formative-themes}

Initial coding produced 140 codes organized into fifteen descriptive groupings. These groupings organized kinds of interview material and were not themselves the final themes. Table~\ref{tab:formative-theme-summary} provides illustrative links between initial codes and the four themes reported in the main text. It is not an exhaustive or mutually exclusive assignment of codes to themes. Initial code identifiers, including F1--F4, are distinct from the F1--F4 theme labels.

Table~\ref{tab:app-formative-cited-codes} gives the definitions of the codes cited in Table~\ref{tab:formative-theme-summary}.

\begin{table}[!htbp]
\centering
\small
\caption{Illustrative connections between formative themes and initial codes. Code definitions are given in Table~\ref{tab:app-formative-cited-codes}.}
\label{tab:formative-theme-summary}
\renewcommand{\tabularxcolumn}[1]{m{#1}}
\begin{tabularx}{\linewidth}{@{}m{0.08\linewidth} Y >{\raggedright\arraybackslash}m{0.22\linewidth}@{}}
\toprule
\textbf{Theme} & \textbf{Supporting aspects of the initial coding} & \textbf{Code IDs} \\
\midrule
F1 & Loss of capture reasons and identifying context, and difficulty distinguishing personally important photographs from routine records. & E1, E2, E3, E11 \\
\rowrule
F2 & Creative or professional purposes, domain knowledge, existing interests, and personal experiences shaping attention. & A4, A8, B2, B6 \\
\rowrule
F3 & Knowledge redirecting attention and comparisons with works encountered across visits. & H13, F3, F7 \\
\rowrule
F4 & Personally meaningful organization, interest in understanding one's taste, and concerns about inferring preference from photographs alone. & J1, J3, J8, K1, K5, K6 \\
\bottomrule
\end{tabularx}
\end{table}

\begin{table}[!htbp]
\centering
\small
\caption{Definitions of the initial codes cited in Table~\ref{tab:formative-theme-summary}.}
\label{tab:app-formative-cited-codes}
\renewcommand{\tabularxcolumn}[1]{m{#1}}
\begin{tabularx}{\linewidth}{@{}>{\raggedright\arraybackslash}m{0.34\linewidth} Y@{}}
\toprule
\textbf{Code} & \textbf{Definition} \\
\midrule
\textbf{A4. Viewing as inspiration sourcing} & Visiting to gather material for creative, academic, or professional activities, including writing, video production, and design. \\
\rowrule
\textbf{A8. Domain expertise shaping attention} & Describing how training in history, art history, aesthetics, or curating shapes the aspects of artworks attended to. \\
\rowrule
\textbf{B2. Subject matter matching a personal interest} & Photographing a work because its subjects or iconography connect with an existing interest, such as mythology, saints, or dogs. \\
\rowrule
\textbf{B6. Personal-biographical resonance} & Connecting a work with personal circumstances or experiences, including illness, medication, employment concerns, or pet ownership. \\
\rowrule
\textbf{E1. Loss of the reason for capture} & Retaining a photograph while being unable to recall why the work was selected for capture. \\
\rowrule
\textbf{E2. Loss of venue or exhibition context} & Being unable to recall the museum or exhibition associated with a photograph. \\
\rowrule
\textbf{E3. Loss of identifying information} & Being unable to recover a work's title or artist, including when a photographed label is illegible. \\
\rowrule
\textbf{E11. Difficulty distinguishing personally important photographs} & Describing photographs of personally meaningful works as difficult to distinguish from routine or precautionary captures. \\
\rowrule
\textbf{F3. Cross-visit comparison} & Returning to earlier photographs to check whether a work was previously encountered or to compare related works or versions. \\
\rowrule
\textbf{F7. Comparison during a later visit} & Needing to retrieve earlier photographs while viewing another exhibition, where searching the camera roll can be inconvenient. \\
\rowrule
\textbf{H13. Knowledge changing what is noticed} & Describing acquired context as drawing attention to features or relationships that had previously gone unnoticed. \\
\rowrule
\textbf{J1. Organization by exhibition or visit date} & Using or preferring an exhibition or visit date as a way to locate and organize memories of works. \\
\rowrule
\textbf{J3. Organization by subject or theme} & Grouping or wanting to group works around shared subjects, themes, or other meaningful categories. \\
\rowrule
\textbf{J8. Personally defined organizing axes} & Wanting categories based on personal experience, emotional intensity, social concerns, or other individually meaningful distinctions. \\
\rowrule
\textbf{K1. Interest in understanding one's taste} & Expressing curiosity about recurring preferences or difficulty articulating what one tends to like. \\
\rowrule
\textbf{K5. Concern about inferring taste from all photographs} & Warning that photographs taken out of obligation or for documentation can misrepresent personal preference. \\
\rowrule
\textbf{K6. Recurring questions as evidence of interest} & Suggesting that repeated questions can indicate interests that photographs alone do not adequately convey. \\
\bottomrule
\end{tabularx}
\end{table}

\subsection{CCO Coverage Analysis}
\label{app:formative-cco}

Separately from the reflexive thematic analysis, two researchers independently coded participant-level coverage across the nine CCO chapters. Ten participants and nine chapters yielded 90 binary participant--chapter judgments. Each positive judgment required a supporting segment identifier.

Eligible evidence included three forms of expression: information wants, capture or attention motives, and organizing preferences. A positive judgment indicated that a participant had expressed attention or interest relevant to a chapter. It did not require an explicit information request or use of CCO terminology. A negative judgment indicated that no eligible supporting statement was identified, not that the participant lacked that interest. Because one participant could contribute evidence to multiple chapters, the coverage counts were not mutually exclusive.

General suggestions about system design were not sufficient on their
own to establish CCO chapter coverage. Accounts of participants' own
practices or standing interests remained eligible regardless of where
they occurred in the interview, including during the discussion of
possible future support. Interests not accommodated by the nine
chapters were recorded separately.

Both coders judged \LensCreator{} and \LensPhysical{} present for all ten participants, and \LensStylistic{} and \LensSubject{} for nine.

\subsection{MyMuEx Trigger Selection}
\label{app:formative-mymuex}

Coding of the formative interview material identified 21 of the 23 experiential triggers defined by MyMuEx. These 21 triggers were adopted as the system's candidate set for experiential annotation.

\section{System Implementation Details}
\label{app:system}

Table~\ref{tab:dependencies} summarizes the inputs and method of each function described in Section~\ref{sec:system} and whether it requires a catalogue match for the photographed artwork.

\begin{table}[!htbp]
  \caption{Inputs of each MyMuse function and whether the function requires a catalogue match for the photographed artwork.}
  \label{tab:dependencies}
  \small
  \renewcommand{\tabularxcolumn}[1]{m{#1}}
  \setlength{\tabcolsep}{4pt}
  \begin{tabularx}{\linewidth}{@{}>{\raggedright\arraybackslash}m{0.44\linewidth} Y >{\raggedright\arraybackslash}m{0.2\linewidth} >{\raggedright\arraybackslash}m{0.13\linewidth}@{}}
    \toprule
    \textbf{Function} & \textbf{Inputs} & \textbf{Method} & \textbf{Catalogue match} \\
    \midrule
    Response generation & Catalogue record; Wikidata artist data; ArtRAG context; previous reply; selected prior records & \texttt{gpt-5.4-mini}, text only & Required for the first three inputs \\
    \rowrule
    Utterance annotation & Visitor utterance & \texttt{gpt-5.4-mini} & Not required \\
    \rowrule
    PMG retrieval & Stored annotations & Shared (aspect, trigger) pairs & Not required \\
    \midrule
    Related works: visual & Visitor's photograph & CLIP ViT-B/32 & Not required \\
    \rowrule
    Related works: artist & Artist name & Corpus match; Met collection search & Required \\
    \rowrule
    Related works: subject & Catalogue subject terms & Controlled vocabulary & Required \\
    \midrule
    Report planning & Annotations and evidence spans & Deterministic planner & Not required \\
    \rowrule
    Report wording & Planned observations & \texttt{gpt-5.4-mini}, checked; deterministic fallback & Not required \\
    \midrule
    Report lens: \LensPhysical{} & Visitor's photograph; annotations & Color and lightness measures & Not required \\
    \rowrule
    Report lens: \LensSubject{} & Catalogue subject terms & Controlled vocabulary & Required \\
    \rowrule
    Report lenses: \LensCreator{}; \LensStylistic{} & Artist; date & Catalogue fields & Required \\
    \bottomrule
  \end{tabularx}
\end{table}

\section{Technical Evaluation: Additional Methods and Results}
\label{app:tech-eval}

This appendix provides implementation and scoring details, prompt structures,
and additional results for the three technical evaluations.

\begingroup
\setlength{\emergencystretch}{2em}
\fvset{
  fontsize=\small,
  baselinestretch=1.08,
  breaklines=true,
  breakanywhere=true,
  frame=single,
  framerule=0.3pt,
  framesep=5pt,
  rulecolor=\color{black!25},
  breaksymbolleft={},
  breaksymbolright={}
}

\subsection{Artpedia Comparison on a Shared Subset}
\label{app:artpedia}
\label{supp:artpedia}

All four configurations were evaluated on the same 310 Artpedia~\cite{stefanini19}
items with recoverable images: ArtRAG~\cite{wang25} and MyMuse, each with
\texttt{gpt-4o-mini} and with \texttt{gpt-5.4-mini}.

\subsubsection{Inputs and Prompting}

All MyMuse configurations received textual metadata and context without
image input and used zero-shot prompting. The ArtRAG configurations
used image input and two-shot prompting. MyMuse additionally received
\texttt{artist}, \texttt{medium}, and \texttt{region} metadata. The MyMuse
task message included the instruction ``strictly under 30 words,'' and the
ArtRAG configurations received the same length instruction.

The \texttt{gpt-5.4-mini} configurations replaced only the model used for
final generation. MyMuse reused the stored system and task messages of its
\texttt{gpt-4o-mini} configuration verbatim, without reassembling the retrieved
context, so the supplied context and routing decisions were the same in both
MyMuse configurations. ArtRAG kept
\texttt{gpt-4o-mini} for its retrieval steps, including image-based reranking,
and used \texttt{gpt-5.4-mini} for the final description call with the same
image input and two-shot prompt. Both systems therefore used
\texttt{gpt-4o-mini} for language-model calls during retrieval, as ArtRAG does
within the deployed MyMuse system, and \texttt{gpt-5.4-mini} only for final
generation.

\subsubsection{Generation and Scoring}

MyMuse's \texttt{gpt-4o-mini} configuration used
\texttt{temperature=1.0}.
Both \texttt{gpt-5.4-mini} configurations used the API default temperature,
and all 310 generations completed for each.{\emergencystretch=8em\par}

A shared evaluator scored all four output sets. Reference-based metrics
used both Artpedia's \texttt{contextual\_sentences} and
\texttt{visual\_sentences}, with PTBTokenizer preprocessing for
BLEU-1--4~\cite{papineni02}, METEOR~\cite{banerjee05}, ROUGE-L~\cite{lin04},
and SPICE~\cite{anderson16}. CLIPScore~\cite{hessel21} used ViT-B/32 to assess image--text
alignment without textual references. Responses were scored at their
original generated lengths, without post-hoc word-count truncation.{\emergencystretch=8em\par}

\subsubsection{Results}

Table~\ref{tab:app-artpedia-shared310} reports the full shared-subset
results and mean response lengths.

\begin{table}[!htbp]
  \centering
  \caption{Performance on the shared 310-item Artpedia subset.
  Metric values are multiplied by 100. Mean response length is reported
  in words. All responses are evaluated at their original generated length.
  The \texttt{gpt-5.4-mini} configurations replace only the model used for
  final generation. Rows are grouped by generation model, and bold marks
  the higher value within each group.}
  \label{tab:app-artpedia-shared310}
  \small
  \setlength{\tabcolsep}{4pt}
  \begin{tabular*}{\linewidth}{@{\extracolsep{\fill}}llrrrrrrrrr@{}}
    \toprule
    \textbf{Model} & \textbf{System}
      & \textbf{BLEU-1} & \textbf{BLEU-2} & \textbf{BLEU-3} & \textbf{BLEU-4}
      & \textbf{METEOR} & \textbf{ROUGE-L} & \textbf{SPICE} & \textbf{CLIPScore} & \textbf{Mean words} \\
    \midrule
    \texttt{gpt-4o-mini} & ArtRAG
      & 35.28 & 17.22 & 8.77 & 4.75 & 15.22 & 22.74 & \textbf{9.35} & \textbf{84.88} & 57.4 \\
    & MyMuse
      & \textbf{45.96} & \textbf{25.31} & \textbf{16.05} & \textbf{10.55} & \textbf{16.38} & \textbf{27.49} & 8.25 & 77.49 & 25.8 \\
    \midrule
    \texttt{gpt-5.4-mini} & ArtRAG
      & 33.94 & 15.61 & 7.83 & 4.21 & 13.74 & 20.28 & 7.87 & \textbf{83.23} & 40.2 \\
    & MyMuse
      & \textbf{44.51} & \textbf{24.04} & \textbf{14.52} & \textbf{9.38} & \textbf{16.30} & \textbf{23.71} & \textbf{8.14} & 77.36 & 23.2 \\
    \bottomrule
  \end{tabular*}
\end{table}

Because MyMuse can use ArtRAG context, we examined whether its higher BLEU-4 and ROUGE-L scores were confined to works covered by ArtRAG's knowledge graph. We compared the two \texttt{gpt-4o-mini} configurations within two predefined coverage strata: 140 covered works and 170 works outside coverage. For MyMuse versus ArtRAG, BLEU-4 was $9.80$ versus $4.46$ on covered works and $11.13$ versus $4.99$ outside coverage. ROUGE-L was $26.95$ versus $22.65$ and $27.93$ versus $22.82$, respectively. The higher scores on these two metrics therefore extended to works outside ArtRAG's knowledge-graph coverage.

\subsection{Annotation Agreement: Methods and Additional Results}
\label{app:annotation-agreement}
\label{supp:annotation-coding}

\subsubsection{Coding Procedure}

Two researchers independently annotated all 332 visitor utterances from
the MyMuse condition of Study~1. Coders received
utterances, artwork titles, artist names, coding instructions, and
published definitions. Model annotations and participant identifiers
were withheld.

The codebook contained 21 visitor-addressable CCO elements and all 23
MyMuEx triggers. Coders assigned the vocabularies independently, with no
limit on labels per utterance, and assigned only labels supported by the
text. Asking a question did not itself imply an experiential trigger.
For uncertain judgments, coders
marked \texttt{uncertain=1} and entered the most plausible code.

Human annotations covered the same 21 MyMuEx triggers adopted by the
system. Neither coder assigned \emph{Contribution} or \emph{VIP}, the two
triggers outside the system vocabulary. The agreement analysis reported
here concerns only the four CCO-derived lenses.

\subsubsection{Model Annotations}

We used annotations stored during the study sessions rather than
regenerating them. Annotation occurred after the visitor-facing response,
using \texttt{gpt-5.4-mini} at \texttt{temperature=0}.

The CCO call received the utterance and 21 candidate elements with their
published definitions. When needed to resolve a reference, the preceding
system reply was supplied as context but was not annotated or quoted as
evidence. The MyMuEx call received the utterance and 21 selected triggers.
A separate call linked CCO elements and MyMuEx triggers only where the
utterance supported the relation. Co-occurrence alone was insufficient.

\subsubsection{Mapping to Lenses}

Table~\ref{tab:app-annotation-mapping} maps each coder's and the model's
CCO assignments to four binary lens labels. A lens was present when at
least one of its elements was assigned. Utterances containing only
elements outside the four lenses received four zero values. The human
reference was the union of the two coders' lens assignments.

\begin{table}[!htbp]
\centering
\small
\caption{Mapping from CCO coding labels to the four lenses.
Identifiers C01--C21 number the 21 CCO elements in the codebook.}
\label{tab:app-annotation-mapping}
\label{tab:supp-annotation-mapping}
% Both columns are vertically centred so that a lens name sits in the middle of its row.
\renewcommand{\tabularxcolumn}[1]{m{#1}}
\begin{tabularx}{\linewidth}{@{}m{0.27\linewidth}Y@{}}
\toprule
\textbf{Lens} & \textbf{CCO elements} \\
\midrule
\LensCreator{}
& C03 Controlled Creator; C04 Creator display. \\
\rowrule
\LensPhysical{}
& C05 Color; C06 Condition; C07 Facture; C08 Inscriptions;
C09 Material; C10 Materials and Techniques display;
C11 Measurements display; C12 Physical Description; C13 Technique. \\
\rowrule
\LensStylistic{}
& C14 Culture; C15 Display Date; C16 Style. \\
\rowrule
\LensSubject{}
& C19 Controlled Subject; C20 Subject display. \\
\rowrule
No lens
& C01 Title; C02 Work Type; C17 Creation Location;
C18 Current Location; C21 Description. \\
\bottomrule
\end{tabularx}
\end{table}

\subsubsection{Agreement Measures and Results}

Table~\ref{tab:app-lens-agreement} reports Cohen's $\kappa$ per lens across
332 utterances and pooled across 1,328 utterance--lens decisions.
Exact-set agreement required matching all four lens labels for an
utterance.

\begin{table}[!htbp]
  \centering
  \caption{Agreement for the four lenses. Agreement percentages
  include matching presence and matching absence. Model--human
  comparisons use the union of the two coders' lens assignments.}
  \label{tab:app-lens-agreement}
  \small
  \begin{tabular}{lrrrr}
    \toprule
    & \multicolumn{2}{c}{\textbf{Inter-coder}}
    & \multicolumn{2}{c}{\textbf{Model--human union}} \\
    \cmidrule(lr){2-3}
    \cmidrule(lr){4-5}
    \textbf{Lens}
      & $\kappa$ & \textbf{Agreement (\%)}
      & $\kappa$ & \textbf{Agreement (\%)} \\
    \midrule
    \LensCreator{} & .767 & 93.7 & .683 & 90.4 \\
    \LensPhysical{} & .839 & 94.0 & .769 & 91.3 \\
    \LensStylistic{} & .891 & 97.3 & .647 & 91.0 \\
    \LensSubject{} & .867 & 94.3 & .662 & 84.9 \\
    \midrule
    Pooled                     & .847 & 94.8 & .701 & 89.4 \\
    \bottomrule
  \end{tabular}
\end{table}

The two sensitivity analyses in Table~\ref{tab:app-annotation-sensitivity}
were applied separately to the full set of 332 utterances. One excluded
20 utterances flagged as uncertain by either coder, leaving 312. The
other excluded 72 utterances for which both the model and human union
assigned no lens, leaving 260.

\begin{table}[!htbp]
  \centering
  \caption{Sensitivity analyses for model--human agreement. Exact-set
  matches are reported as counts and percentages of utterances in each
  subset. Binary agreement is calculated across four lens decisions per
  utterance.}
  \label{tab:app-annotation-sensitivity}
  \small
  \setlength{\tabcolsep}{5pt}
  \begin{tabular}{lrrrr}
    \toprule
    \textbf{Subset}
      & \textbf{Utterances}
      & \textbf{Pooled $\kappa$}
      & \shortstack{\textbf{Binary agreement}\\\textbf{(\%)}}
      & \shortstack{\textbf{Exact-set matches}\\\textbf{$n$ (\%)}} \\
    \midrule
    All utterances
      & 332 & .701 & 89.4 & 214 (64.5) \\
    Excluding uncertain utterances
      & 312 & .700 & 89.0 & 198 (63.5) \\
    Excluding jointly empty assignments
      & 260 & .674 & 86.4 & 142 (54.6) \\
    \bottomrule
  \end{tabular}
\end{table}

\subsection{Report Composition: Methods and Additional Results}
\label{app:report-composition}
\label{supp:report-composition}

\subsubsection{Inputs, Prompts, and Output Fields}

The three conditions shared one contract per participant, which contained
the participant's records and artwork metadata.

The A/B prompt omitted C's instruction prohibiting new interpretations
and treating composition decisions as already made, because A and B
selected the composition. The prompt structures are provided in
Sections~\ref{app:report-prompt-ab} and~\ref{app:report-prompt-c}.

To compare C with A and B on the same fields, C's entry lens, anchor
artwork, mentioned artworks, and claimed patterns were read from the
planner's output rather than extracted from generated prose, and its
narrative text consisted of the accepted generated sentences in display
order.

\subsubsection{Generation and Validation}

All conditions used \texttt{gpt-5.4-mini} with a ceiling of 4,000 completion
tokens per call.
The default setting omitted the temperature field, and the second setting
requested \texttt{temperature=0}, which the API accepted for every call.

C reused the production planning, fact extraction, prompt, sentence
validation, and merge functions while bypassing the cache. It made up to
two generation calls, retaining accepted sentences from the first call and
filling remaining slots from the second.

B began with A's output for the same participant, repetition, and
temperature setting. Its batch validator checked narrative sentences for
unsupported numbers and names, speculative
expressions, and fixed-trait descriptions, and removed invalid entry-lens
values and artwork identifiers. Pattern phrases were checked only for
unsupported numbers and names.

In B, rejection of a narrative sentence or structural field triggered at
most one additional query with the same prompt and contract. Both responses were
validated. Each nonempty first-pass field was retained, with the second
pass filling only empty fields. Individual rejected sentences within a
nonempty \texttt{narrative\_text} field were not replenished, and pattern
filtering alone did not trigger re-querying.

\subsubsection{Stability Measures}
\label{app:report-stability-methods}

Within each participant, condition, and temperature setting, entry-lens
and anchor-artwork disagreement were computed as
\[
D = 1-\frac{\max_v n_v}{m},
\]
where $m$ is the number of retained runs and $n_v$ the number assigning
value $v$. Missing values were represented by an empty string. Artwork-
and pattern-set stability were the mean Jaccard similarity over all
$\binom{m}{2}$ repetition pairs, with
\[
J(S_i,S_j)=\frac{|S_i\cap S_j|}{|S_i\cup S_j|},
\qquad J(\varnothing,\varnothing)=1.
\]
Participant-level values were summarized by their mean and sample
standard deviation. Pairwise measures were undefined for fewer than two
retained runs.

Three of the 320 A runs returned unparseable JSON and were left out together
with the corresponding B runs. C had no such runs, because its structural
fields come from its plan.

Pattern normalization matched complete CCO chapter names, then CCO
elements (longest matching label first), MyMuEx trigger labels or codes,
and a fixed alias table. CCO matches mapped to chapters. MyMuEx
matches generally retained trigger-specific keys, with \texttt{curation}
and \texttt{atmosphere} mapped to Context. Unmatched phrases retained
separate keys after Unicode, punctuation, case, and rule-based
suffix normalization. Pattern similarity therefore combines chapter-level,
trigger-level, and normalized-string matches.

\subsubsection{Results with Temperature Zero Requested}
\label{app:eval-temp}

Table~\ref{tab:app-stability-temp0} reports the second temperature setting
using the measures defined in Section~\ref{app:report-stability-methods}.
The same sixteen participant records and ten repetitions per condition
were used, with \texttt{temperature=0} requested.

\begin{table}[H]
  \centering
  \caption{Report-composition stability with \texttt{temperature=0}
  requested. Values are means over sixteen participants, with sample
  standard deviations in parentheses. Jaccard values summarize pairwise
  set similarities across retained repetitions within each participant.}
  \label{tab:app-stability-temp0}
  \small
  \begin{tabular}{lccc}
    \toprule
    \textbf{Measure} & \textbf{A} & \textbf{B} & \textbf{C} \\
    \midrule
    Entry-lens disagreement
      & 0.056 (0.126) & 0.056 (0.126) & 0.000 \\
    Anchor-artwork disagreement
      & 0.075 (0.118) & 0.075 (0.118) & 0.000 \\
    Artwork-set Jaccard
      & 0.925 (0.112) & 0.925 (0.112) & 1.000 \\
    Pattern-set Jaccard
      & 0.314 (0.132) & 0.313 (0.132) & 1.000 \\
    \bottomrule
  \end{tabular}
\end{table}

\subsubsection{A/B Prompt Structure}
\label{app:report-prompt-ab}
\label{supp:report-prompt-ab}

A's initial generation and B's conditional re-query used the shared contract
JSON as the user message.

\begin{Verbatim}
You write a collection-level reflection report from museum visit records.

The JSON contains one person's artwork catalogue information, verbatim
utterances, CCO element annotations and evidence spans, MyMuEx experiential
triggers, relationships between CCO elements and triggers, contextual
records not assigned to an artwork, and provenance information.

Decide the following:
1. Entry lens: Choose Creator, Physical, Stylistic, or Subject (CCO
   Chapters 2, 3, 4, and 6, respectively). Choose Context if the records
   primarily concern placement or atmosphere outside the artworks, or
   Other if no lens is supported.
2. Anchor artwork: Select a starting artwork with actual evidence
   supporting the selected lens.
3. Recurring patterns: Identify interests supported by at least two
   different artworks. Repeated mentions within one artwork count as
   one supporting artwork.
4. Report prose: Begin with the anchor, proceed through the patterns,
   and conclude the report. Include only sections supported by
   evidence; keep the report short when evidence is limited.

Constraints:
- Do not state facts absent from the JSON. Use only artwork names,
  element names, trigger names, and numbers present in the JSON.
- Do not characterize the person's disposition, taste, or personality.
  State only what recurred in the records.
- Do not use speculative expressions such as "perhaps," "seems,"
  "appears," or "feels."
- Write for the person who made the records; do not refer to them in
  the third person as "the participant," "the user," or "this person."

Return only one JSON object, without a code fence:
{"entry_lens": "Creator | Physical | Stylistic | Subject | Context | Other",
 "anchor_artwork": "<artwork_id -- copy the value from artworks in the JSON>",
 "artworks_mentioned": ["<artwork_id>", ...],
 "patterns_claimed": ["<short phrase>", ...],
 "narrative_text": "<complete report prose>"}
\end{Verbatim}

\subsubsection{C Prompt Structure}
\label{app:report-prompt-c}
\label{supp:report-prompt-c}

C received the plan-specific facts JSON as the user message.

\begin{Verbatim}
You edit sentences for a report on museum visit records. The analysis
and composition decisions are already complete. Express each supplied
result as one sentence.

Constraints:
- Do not state facts absent from the JSON. Use only artwork names,
  element names, trigger names, and numbers present in the JSON.
- Do not characterize the person's disposition, taste, or personality.
  State only what recurred in the records.
- Do not create new interpretations. Which lens is important and what
  each result means have already been determined.
- Do not use speculative expressions such as "perhaps," "seems,"
  "appears," or "feels."
- Write for the person who made the records; do not refer to them in
  the third person as "the participant," "the user," or "this person."
- Each sentence must contain no more than 120 characters, including
  spaces.

What to write:
- anchor_reason: Explain why the already-selected artwork provides a
  starting point for the selected lens, using only anchor.signal and
  the actual utterances about that artwork. Connect anchor.signal.why
  to the quoted observation. Paraphrase what drew attention without
  inferring intentions or making additional judgments; do not merely
  restate annotation labels. For title_vs_subject, describe the
  mismatch between the title's referent and the visible subject.
  For question_answered, describe the question and the statement that
  followed it. Do not invent reasons absent from the signal. If
  anchor.signal is null, set anchor_reason to null.
- lens_takeaway: Write one sentence per lens present in lenses,
  describing what its elements repeatedly identified across artworks.
  Do not add other lenses.
- final_synthesis: Describe which experiential triggers accompanied
  recurring interests, using only links and reactions.
- term_contrast: Write one sentence per subject term present in terms,
  describing how artworks assigned that term treat the subject
  differently. Use only each artwork's described field (the catalogue's
  description of the image). Specify what differs; do not add subject
  terms or claims absent from these fields.
- record_interest: Write one sentence only when record is present,
  using only record.said. Describe what was asked about that single
  subject and how the inquiry developed, beyond noting that the term
  occurred. Do not include other subjects, other lenses, or an overall
  summary, and do not write names or numbers outside record.term.

Return only one JSON object:
{"anchor_reason": "...", "lens_takeaway": {"subject": "..."},
 "final_synthesis": "...",
 "term_contrast": {"<subject_term>": "...", ...},
 "record_interest": "..."}
Leave a slot null if there is insufficient evidence to write it.
\end{Verbatim}

\endgroup

\section{Additional User Study Results and Measures}
\label{app:user-study}

\subsection{Additional Statistical Results}
\label{app:user-study-statistics}

Tables~\ref{tab:app-study1-stats}--\ref{tab:app-study2-stats} report the full statistics for the results summarized in Section~\ref{sec:results-study1} and Section~\ref{sec:results-study2}. $W$ denotes the Wilcoxon signed-rank statistic, $p$ the unadjusted value, $p_{\mathrm{Holm}}$ the Holm-adjusted value, and $r$ the effect size $|Z|/\sqrt{N}$.

\begin{table}[!htbp]
\centering
\small
\caption{Statistics for the six shared Study~1 outcomes ($N = 16$). Holm correction was applied across the six outcomes.}
\label{tab:app-study1-stats}
\begin{tabular}{lcccccc}
\toprule
\textbf{Outcome} & \textbf{MyMuse} $M$ ($SD$) & \textbf{Baseline} $M$ ($SD$) & $W$ & $p$ & $p_{\mathrm{Holm}}$ & $r$ \\
\midrule
Re-experiencing & 6.00 (1.10) & 5.31 (1.14) & 16.0 & .065 & .137 & .46 \\
Elaboration & 6.50 (0.73) & 6.00 (0.73) & 9.0 & .046 & .137 & .50 \\
Connections among responses & 5.75 (1.44) & 5.00 (1.55) & 7.0 & .032 & .129 & .54 \\
Answer relevance & 6.12 (0.89) & 5.31 (0.87) & 0.0 & .006 & .035 & .69 \\
Engagement & 6.69 (0.48) & 6.06 (0.93) & 0.0 & .026 & .128 & .56 \\
Intention to use & 6.50 (0.89) & 5.81 (1.28) & 8.5 & .093 & .137 & .42 \\
\bottomrule
\end{tabular}
\end{table}

\begin{table}[!htbp]
\centering
\small
\caption{Statistics for the Study~1 direct-comparison items, tested against the scale midpoint of 4 ($1 = $ MyMuse, $7 = $ Baseline). Holm correction was applied across the three items.}
\label{tab:app-study1-direct}
\begin{tabular}{lccccccc}
\toprule
\textbf{Outcome} & \textbf{MyMuse / same / Baseline} & $Mdn$ & $SD$ & $W$ & $p$ & $p_{\mathrm{Holm}}$ & $r$ \\
\midrule
Connections among responses & 12 / 2 / 2 & 2 & 1.49 & 5.0 & .002 & .005 & .76 \\
Re-experiencing & 13 / 1 / 2 & 2 & 1.41 & 4.0 & .001 & .004 & .81 \\
Overall preference & 12 / 0 / 4 & 1.5 & 2.27 & 31.0 & .049 & .049 & .49 \\
\bottomrule
\end{tabular}
\end{table}

After Holm correction, answer relevance was the only shared Study~1 outcome with a statistically significant difference. All three direct-comparison items remained statistically significant. Findability, analyzed separately as a usability check, did not differ significantly (MyMuse $M = 6.31$, $SD = 0.95$; baseline $M = 5.81$, $SD = 1.11$; $p = .145$, $r = .36$). On the MyMuse-specific diagnostic items, personalized Q\&A and spatial reconstruction were each rated $M = 6.06$, with 15 and 14 of 16 participants, respectively, rating them 5 or above.

\begin{table}[!htbp]
\centering
\small
\caption{Statistics for the Study~2 measures ($N = 16$). Holm correction was applied across the three collection-level reflective outcomes. Perceived fidelity was assessed separately as a grounding check.}
\label{tab:app-study2-stats}
\begin{tabular}{lccccc}
\toprule
\textbf{Outcome} & \textbf{Narrative} $M$ ($SD$) & \textbf{Structured} $M$ ($SD$) & \textbf{Narrative higher / equal / Structured higher} & $p$ ($p_{\mathrm{Holm}}$) & $r$ \\
\midrule
Recurring-interest awareness & 6.62 (0.62) & 4.62 (1.71) & 12 / 3 / 1 & .002 (.005) & .76 \\
Cross-artwork connection awareness & 6.62 (0.50) & 4.19 (1.87) & 14 / 2 / 0 & $<$.001 (.003) & .83 \\
Unexpected discovery & 6.31 (1.14) & 3.88 (1.93) & 14 / 1 / 1 & .004 (.005) & .72 \\
Perceived fidelity & 6.69 (0.60) & 6.31 (0.79) & 8 / 7 / 1 & .083 & .43 \\
\bottomrule
\end{tabular}
\end{table}

\subsection{Questionnaire Items}
\label{app:user-study-questionnaires}

Tables~\ref{tab:app-study1-items}--\ref{tab:app-study2-items} list the questionnaire items. Unless otherwise specified, responses used 7-point agreement scales with $1=\mathrm{strongly\ disagree}$, $4=\mathrm{neutral}$, and $7=\mathrm{strongly\ agree}$.

\begin{table}[!htbp]
\centering
\small
\caption{Study~1 questionnaire items. The six shared outcomes and findability were administered after each condition. The final two diagnostic items were administered only after the MyMuse condition.}
\label{tab:app-study1-items}
\renewcommand{\tabularxcolumn}[1]{m{#1}}
\begin{tabularx}{\linewidth}{@{}>{\raggedright\arraybackslash}m{0.25\linewidth} Y@{}}
\toprule
\textbf{Measure} & \textbf{Questionnaire item} \\
\midrule
Re-experiencing & While using this system, my museum viewing experience at the time came vividly to mind. \\
\rowrule
Elaboration & This system helped me develop my thoughts about the artwork more concretely. \\
\rowrule
Connections among responses & This system helped me notice commonalities or connections among my responses to different artworks. \\
\rowrule
Answer relevance & The system's answers matched what I wanted to know. \\
\rowrule
Engagement & While using this system, I wanted to further explore the artworks and my records. \\
\rowrule
Intention to use & I would like to use this system again after an actual museum visit. \\
\rowrule
Findability & In this system, I could easily find and inspect the artworks or records I wanted. \\
\midrule
Personalized Q\&A & It was useful to have my previous questions and reflections reflected in subsequent responses. \\
\rowrule
Spatial reconstruction & It was useful to have the artworks and records presented in a spatially reconfigured form. \\
\bottomrule
\end{tabularx}
\end{table}

\begin{table}[!htbp]
\centering
\small
\caption{Study~1 direct-comparison items. All three items used a bipolar 7-point scale with $1=\mathrm{MyMuse}$, $4=\mathrm{about\ the\ same}$, and $7=\mathrm{Baseline}$.}
\label{tab:app-study1-direct-items}
\renewcommand{\tabularxcolumn}[1]{m{#1}}
\begin{tabularx}{\linewidth}{@{}>{\raggedright\arraybackslash}m{0.25\linewidth} Y@{}}
\toprule
\textbf{Measure} & \textbf{Questionnaire item} \\
\midrule
Connections among responses & Which system was more helpful for noticing commonalities or connections among your responses to different artworks? \\
\rowrule
Re-experiencing & Which system was more helpful for vividly recalling your museum viewing experience at the time? \\
\rowrule
Overall preference & If you were to choose one system for revisiting your museum experience after a museum visit, which system would you prefer to use? \\
\bottomrule
\end{tabularx}
\end{table}

\begin{table}[!htbp]
\centering
\small
\caption{Study~2 questionnaire items, administered after both the Narrative Report and the Structured Record.}
\label{tab:app-study2-items}
\renewcommand{\tabularxcolumn}[1]{m{#1}}
\begin{tabularx}{\linewidth}{@{}>{\raggedright\arraybackslash}m{0.25\linewidth} Y@{}}
\toprule
\textbf{Measure} & \textbf{Questionnaire item} \\
\midrule
Recurring-interest awareness & It helped me notice interests that appeared repeatedly across multiple artworks. \\
\rowrule
Cross-artwork connection awareness & It helped me notice connections among my experiences or reactions to different artworks. \\
\rowrule
Unexpected discovery & I discovered interests or connections that I had not previously considered. \\
\rowrule
Perceived fidelity & I felt that the presented content accurately reflected the records I had actually left. \\
\bottomrule
\end{tabularx}
\end{table}

\subsection{Example Exchange from Study 1}
\label{app:example-exchange}

% Keep this short paragraph on one page.
{\interlinepenalty=10000 Table~\ref{tab:example-exchange} reproduces P09's exchange about Pablo Picasso's \emph{Weeping Woman} (1937; Figure~\ref{fig:study-cases}). The notes in the work's catalogue record summarized the holding museum's description of the work, and the work met the ArtRAG routing conditions (Section~\ref{sec:grounded-inquiry}). When the record did not describe how the work reached its current collection, the response said so. The right column lists the context sources logged for each response.\par}

\begin{table}[!htbp]
\centering
\small
\caption{P09's exchange about \emph{Weeping Woman} in the MyMuse condition of Study~1. Bracketed ellipses mark omitted passages.}
\label{tab:example-exchange}
\begin{tabularx}{\linewidth}{@{}p{0.07\linewidth} Y >{\raggedright\arraybackslash}p{0.2\linewidth}@{}}
\toprule
\textbf{Turn} & \textbf{Text} & \textbf{Context sources} \\
\midrule
P09 & This work is by Picasso, whom I really like. Could you tell me what you know about the personal and historical influences on the artist when he made it? & \\
\rowrule
MyMuse & Of course. \emph{Weeping Woman} was made in 1937, and Picasso's personal life and the historical situation of that time overlap strongly in it. The largest historical background is the Spanish Civil War and the 1937 bombing of Guernica. That same year, Picasso addressed the suffering of civilians in \emph{Guernica}, and \emph{Weeping Woman} can be seen as a follow-up image that grew out of that shock. [\ldots] On the personal side, it matters that the model was Dora Maar, Picasso's partner at the time. The NGV's record also introduces the work in connection with their complex and intense relationship. Rather than asserting that the painting directly depicts that relationship, however, it is more accurate to say that Picasso's personal tensions and emotions permeate the work. [\ldots] & Catalogue record\newline Wikidata\newline ArtRAG\newline Model knowledge \\
\midrule
P09 & By the way, how did this work end up in Australia? & \\
\rowrule
MyMuse & Good question. However, the NGV's record does not describe the specific provenance or the route by which this work came to Australia, so I cannot tell you exactly whose hands it passed through. What is certain is that it is an oil painting Pablo Picasso made in 1937 and that it is now in the NGV's collection. Modern works like this usually enter overseas museums through private collectors, dealers, galleries, gifts, or purchases, but for this work those details cannot be confirmed from the records provided. If you like, I can outline the routes by which it most likely came to Australia, based on how the NGV has generally acquired works by Picasso. & Catalogue record\newline Wikidata\newline ArtRAG\newline Model knowledge\newline Previous reply\newline Prior visitor records \\
\midrule
P09 & Then what influence did Picasso have on the Cubist movement in art history? I'm curious about Picasso's artistic world. & \\
\rowrule
MyMuse & Picasso was a key figure who effectively co-created Cubism, so his influence on art history is very large. In particular, breaking objects down and reassembling them as if seen from several angles at once, rather than from a single viewpoint, changed painting from depicting what is seen to perceiving the object anew. [\ldots] In works such as \emph{Weeping Woman}, Cubist fragmentation goes beyond formal experiment and is used to express emotions such as grief and pain more strongly. For Picasso, Cubism was not a cold technique but a powerful means of condensing human inner life and the tragedies of his time. & Catalogue record\newline Wikidata\newline ArtRAG\newline Model knowledge\newline Previous reply\newline Prior visitor records \\
\bottomrule
\end{tabularx}
\end{table}

\end{document}